\documentclass[fleqn,usenatbib,useAMS]{mnras}

\usepackage{graphicx}	
\usepackage{amsmath}	
\usepackage{subcaption}
\usepackage{float}
\usepackage{cleveref}
\usepackage{multirow}
\usepackage{multicol}
\usepackage{bm}		
\usepackage{pdflscape}
\usepackage{longtable}
\usepackage{amssymb}	

\usepackage[T1]{fontenc}
\usepackage{ae,aecompl}

\usepackage{newtxtext,newtxmath}

\title[Updated Bounds on ALPs]{Updated Bounds on Axion-Like Particles from X-ray Observations}

\author[S. Schallmoser et al.]{
Simon Schallmoser,$^{1}$\thanks{E-mail: schallmoser@lmu.de}
Sven Krippendorf,$^{2}$
Francesca Chadha-Day$^{3}$
and Jochen Weller$^{1,4}$
\\
$^{1}$Universitäts-Sternwarte, Fakultät für Physik, Ludwig-Maximilians Universität München, Scheinerstr.~1, 81679 München, Germany\\
$^{2}$Arnold Sommerfeld Center for Theoretical Physics, Ludwig-Maximilians Universität München, Theresienstr.~37, 80333 M\"unchen, Germany\\
$^{3}$Department of Physics, University of Durham, South Rd, Durham DH1 3LE, United Kingdom\\
$^{4}$Max Planck Institute for Extraterrestrial Physics, Giessenbachstr.~1, 85748 Garching, Germany
}

\date{}

\pubyear{2022}

\begin{document}
\label{firstpage}
\pagerange{\pageref{firstpage}--\pageref{lastpage}}
\maketitle

\begin{abstract}
In this work we revisit five different point sources within or behind galaxy clusters in order to constrain the coupling constant between axion-like particles (ALPs) and photons. We use three distinct machine learning (ML) techniques and compare our results with a standard $\chi^2$ analysis. For the first time we apply approximate Bayesian computation to searches for ALPs and find consistently good performance across ML classifiers. Further, we apply more realistic 3D magnetic field simulations of galaxy clusters and compare our results with previously used 1D simulations. We find constraints on the ALP-photon coupling at the level of state-of-the-art bounds with $g_{a\gamma\gamma} \lesssim 0.6 \times 10^{-12} \, \mbox{GeV}^{-1}$, hence improving on previous constraints obtained from the same observations.
\end{abstract}


\section{Introduction}
Axion-like particles (ALPs) arise within various high-energy physics extensions of the Standard Model of particle physics. In particular, they are guaranteed to appear within supersymmetric extensions with field-dependent couplings, which includes supersymmetric string compactifications (cf.~for example~\citet{st1,st2,st3}). Thus, they provide a well-motivated extension to the Standard Model of particle physics. They couple to photons via the following Lagrangian:
\begin{equation}\label{action}
    \mathcal{L} =\frac{1}{2}\partial_{\mu}a\partial^{\mu}a-\frac{1}{2}m_a^2a^2+g_{a\gamma\gamma}~a~\textbf{E}\cdot\textbf{B} \, ,
\end{equation}
where $a$ denotes the ALP field, $m_a$ the ALP mass, $g_{a\gamma\gamma}$ the coupling constant of ALPs and photons and $\textbf{E}$/$\textbf{B}$ the electric/magnetic field. In a background magnetic field ALPs and photons can interconvert into each other~\citep{mix}. This is the basis of most ALP searches across different ALP masses and couplings~\citep{pdg}. X-ray observations of bright point sources in or behind galaxy clusters have proven to be very useful environments to constrain ALPs~\citep{hydra,ngc1275,non-obs,m87,grating,probe_planck,probe_cmb}.
\begin{figure}
\centering
\includegraphics[width=1\linewidth]{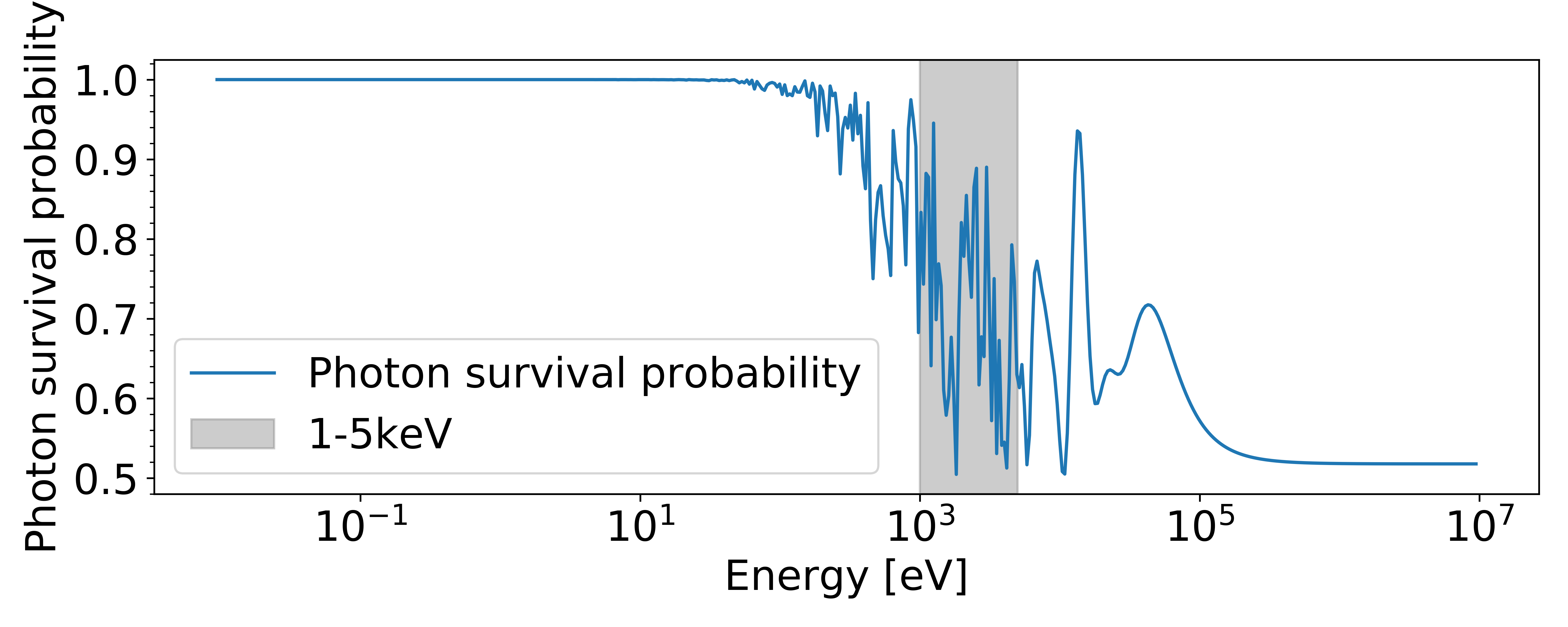}
\caption{Simulated photon survival probability for the Sy1 galaxy 2E3140 within A1795 for a large energy range. The grey shaded area shows the energy range that we have considered for this source. 
In this simulation we use an ALP-photon coupling of $g_{a\gamma\gamma}=5\times 10^{-12} \, \mbox{GeV}^{-1}$.}
\label{fig:surv_prob}
\end{figure}
The main reason for this sensitivity is that galaxy clusters are usually entirely permeated by a magnetic field. Thus, the spectra of point sources, such as active galactic nuclei (AGN) or quasars shining through the cluster would be altered if ALPs exist. As reviewed below, these spectral changes correspond to modulations in the X-ray and $\gamma$-ray regime (cf.~Figure~\ref{fig:surv_prob} for an example visulisation) and hence render X-ray observations of such point sources a very valuable observational window for ALP searches. 

Concretely, spectral modulations arise from photons which are converted to ALPs in the magnetic field of a galaxy cluster which in a first approximation can be modelled as a series of domains, each with a different constant magnetic field and electron density (cf.~Section~\ref{sec:mag_fields} for a detailed discussion of the magnetic field models). In such a domain we can calculate the photon survival probability, i.e.~that a photon remains in a photon state, analytically~\citep{mix}:
\begin{equation}\label{eom}
    P_{\gamma\rightarrow\gamma}=1-\frac{\Theta^2}{1+\Theta^2}\sin^2(\Delta \sqrt{1+\Theta^2}) \, ,
\end{equation}
where $\Theta=2B_{0,n}g_{a\gamma\gamma}\omega / m_{\rm eff}^2$, $\Delta=m_{\rm eff}^2L / (4\omega)$ and $m_{\rm eff}^2=m_a^2-\omega_{\rm pl}^2$. In this work we are interested in constraining ALPs with a mass smaller than the effective photon mass in astrophysical plasmas, i.e.~$m_a \lesssim 10^{-12} \, \mbox{eV}$ which allows us to treat ALPs as massless. $B_{0,n}$ is the magnetic field perpendicular to the direction of travel of the photon, $\omega$ the photon frequency, $\omega_{\rm pl}=\sqrt{4\pi\alpha n_e / m_e}$ the plasma frequency, $\alpha$ the fine-structure constant, $n_e$ the electron density, $m_e$ the electron mass and $L$ the length of the domain.

As our knowledge about the cluster magnetic field is limited, it is approximated by a statistical turbulent magnetic field model which can be constrained by Faraday rotation measures~\citep{Coma} (cf.~Section~\ref{sec:mag_fields}). To obtain information about the expected spectral modulations due to ALPs, a large sample of magnetic fields drawn from the relevant statistical distribution is used. For such a random magnetic field sample, we calculate the survival probability where the initial state of a domain is given by the final state of the previous one. An illustrative example of a photon survival probability is given in Figure~\ref{fig:surv_prob} for one of the sources we discuss in this article, the type I Seyfert galaxy 2E3140 within A1795. The overall amplitude of the oscillations depends on the strength of the coupling and on the magnetic field $B_{0,n}$ -- larger values of $g_{a\gamma\gamma}$ and $B_{0,n}$ lead to larger oscillations. The position of the oscillations differs among magnetic field configurations.

As this estimate of the ALP-signal depends on how realistic the magnetic field of the galaxy cluster is, we consider extensively for the first time more realistic 3D magnetic field simulations in order to constrain ALPs with X-rays.\footnote{For ALP searches in $\gamma$-ray spectra, 3D magnetic field models have been used previously in~\citet{fermi-lat}.} Although they are computationally more involved than 1D magnetic field models, they are more realistic and theoretically better motivated as the underlying magnetic field is divergence free. In order to compare the differences between the two models we apply the 3D model to five sources that have been already investigated with 1D simulations~\citep{non-obs,Day:2019ucy}. We also introduce a re-scaled 1D model which resembles more features of the 3D model (e.g.~the mean strength of the magnetic field).

\begin{figure*}
\centering
\begin{subfigure}{0.5\textwidth}
\includegraphics[width=\textwidth]{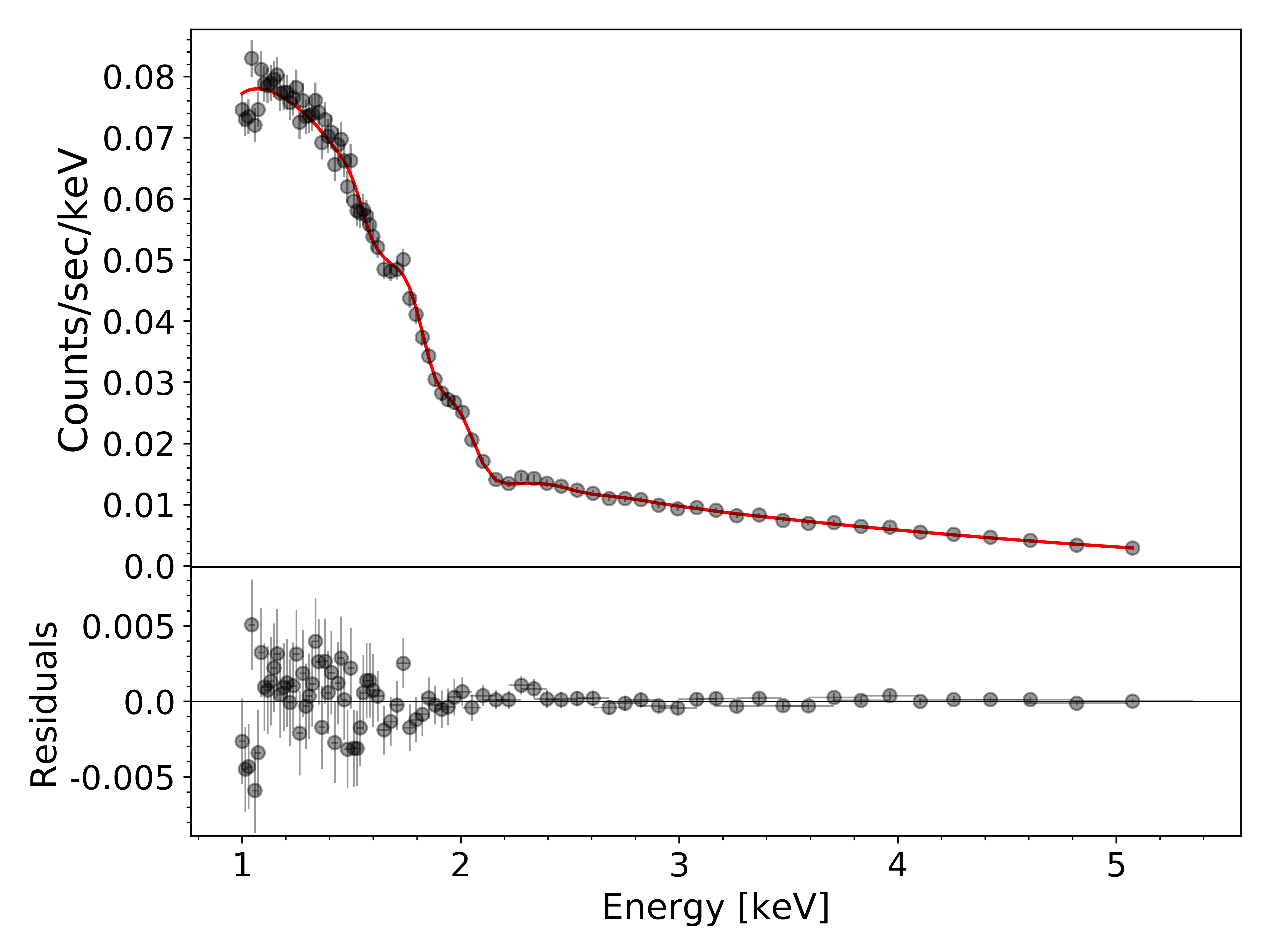}
\end{subfigure}\hspace{-0.25cm}
\begin{subfigure}{0.5\textwidth}
\includegraphics[width=\textwidth]{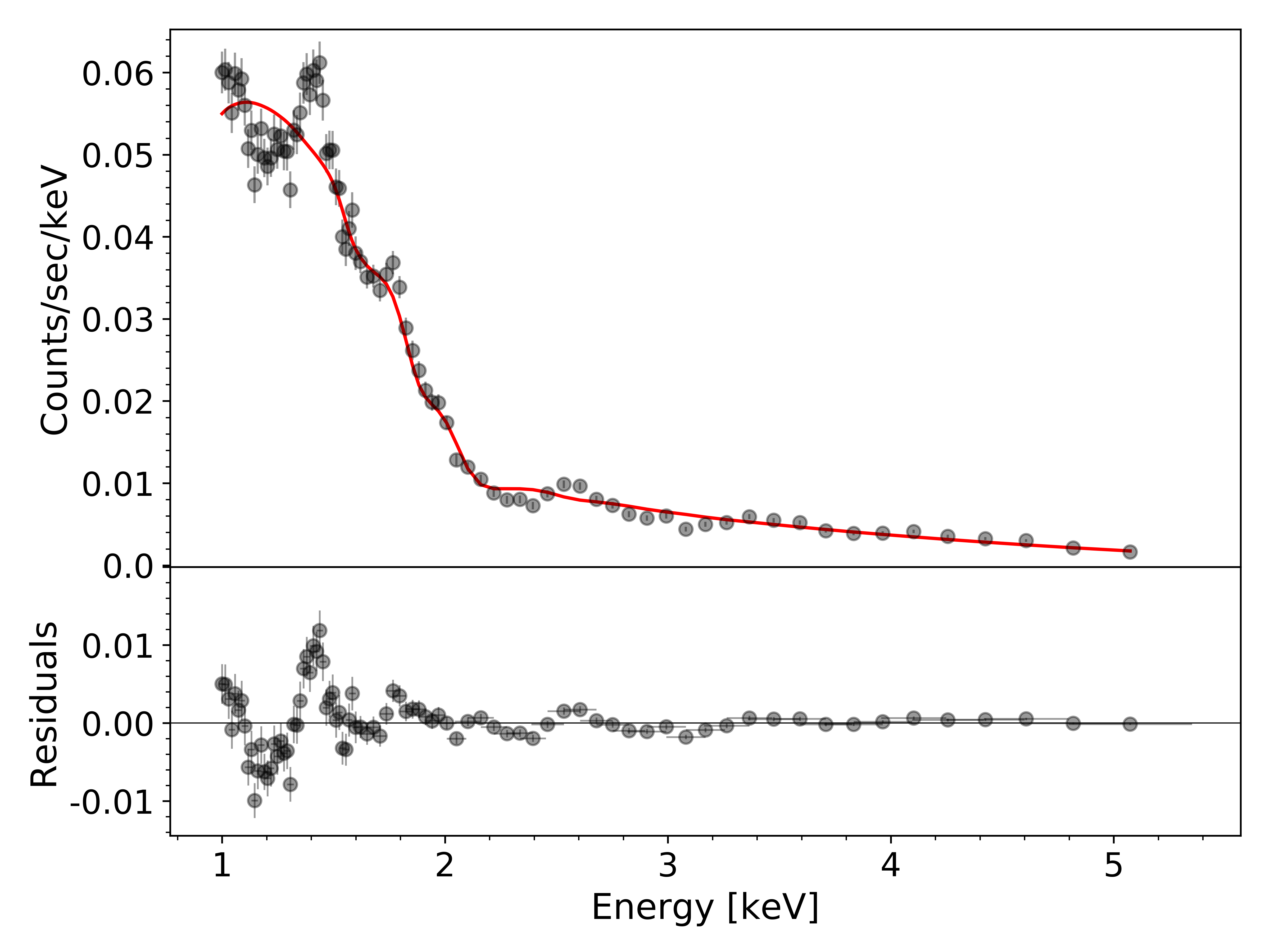}
\end{subfigure}
\caption{\textit{Left:} Spectrum of the Sy1 galaxy 2E3140 within A1795 as observed with \textit{Chandra}. \textit{Right:} Simulated fake spectrum produced by \textit{Sherpa} where we have assumed an ALP-photon coupling of $g_{a\gamma\gamma}=5\times 10^{-12} \, \mbox{GeV}^{-1}$. Both spectra were fitted with a power law and absorption from neutral hydrogen.}
\label{fig:sherpa_comparison}
\end{figure*}

In particular, we use observations of these sources with the \textit{Chandra} X-ray telescope (ACIS instrument). These point source spectra can be well fitted with a power-law and absorption from neutral hydrogen. For instance, the fit of the spectrum of the Sy1 2E3140 galaxy within A1795 is shown on the left of Figure~\ref{fig:sherpa_comparison}.
 In order to constrain ALPs, we need to generate fake spectra including the effects of ALPs. These are generated from the product of this fitted point source spectrum and a sampled photon-survival probability for a certain value of $g_{a\gamma\gamma}$. We then obtain fake spectra using \textit{Sherpa} (version 4.12)~\citep{sherpa}, which adds Poisson noise and instrumental effects such as the detector energy resolution. An example of such a fake spectrum for the Sy1 2E3140 galaxy within A1795 is plotted on the right of Figure~\ref{fig:sherpa_comparison} where we have used the photon survival probability of Figure~\ref{fig:surv_prob}. Overall, fake spectra with ALPs show three distinct features with respect to spectra without ALPs: The former have an overall lower flux and oscillations which increase in their wavelength and intensity with increasing energy. 

From such fake spectra with ALPs constraints on the coupling to photons $g_{a\gamma\gamma}$ can be obtained by comparing the fits of multiple fake spectra -- accounting for the intrinsic Poisson noise and our uncertainty of the cluster magnetic field -- using a $\chi^2$ statistic~\citep{hydra,m87,non-obs}. This procedure does not directly take into account the intrinsic oscillatory features of the ALP signals. Different data representations such as analysing spectra in Fourier space and machine learning (ML) techniques have been shown to increase sensitivity to ALPs~\citep{stat-impr,Day:2019ucy,Marsh:2021ajy}.

In this work we improve these ML methods with a more detailed hyperparameter search and apply approximate Bayesian computation (ApBC)~\citep{apbc1,apbc2} for the first time on ALP searches. We find that the latter leads to more stable bounds across ML approaches. 

This paper is organised as follows. Section~\ref{sec:mag_fields} discusses our models of magnetic fields in galaxy clusters. Section~\ref{sec:astro_systems} provides an overview of the astrophysical sources which we use in this article. In Section~\ref{sec:methods} we present our different bounds methods and in Section~\ref{sec:results} we present their results. Finally, we conclude and give an outlook in Section~\ref{sec:conclusion}. 

\section{Magnetic Field Models of Galaxy Clusters}\label{sec:mag_fields}

In the presence of ALPs, the survival probability of a photon is highly dependent on the magnetic field through which it propagates (cf.~Eq.~\eqref{eom}). The most thorough analysis for magnetic field determination has been performed for the Coma cluster where numerical simulations were compared to Faraday Rotation Measure images~\citep{Coma}. These simulations assumed a turbulent, three-dimensional magnetic field model which is initialized randomly in Fourier space~\citep{murgia}. As these 3D simulations are computationally more expensive and time consuming than 1D approximations, in most previous works simpler 1D magnetic field models along the line of sight have been used to constrain ALPs.

After outlining the 3D model, we compare both approaches and discuss characteristic differences between both approaches. We also introduce an appropriately re-scaled 1D model which can mimic some of the characteristics of the 3D model.

\subsection{3D model}
We are interested in simulating a magnetic field following an inverse power-law power spectrum with a radial profile following the gas distribution. Our simulation follows the approach presented in~\citet{murgia} which proceeds as follows: We first simulate our 3D magnetic field on a lattice with size $2000^3$ which corresponds to a resolution of $1 \, \mbox{kpc}$ when the cluster has a radius of $1\, \mbox{Mpc}$. To generate a divergence-free magnetic field (i.e.~$\nabla \cdot \textbf{B} =0$) with an appropriate power spectrum we start in Fourier space by randomly generating a vector potential with the following power spectrum:
\begin{equation}
    |\tilde{A}_k|^2\sim k^{-(n+2)} \, ,
\end{equation}
where $k$ is only non-zero within a range of $k_{\rm min}=2\pi/\Lambda_{\rm max}$ and $k_{\rm max}=2\pi/\Lambda_{\rm min}$. $\Lambda_{\rm min,max}$ denote the scales over which the magnetic field fluctuates, i.e.~it defines the length scale on which it completely changes its direction. The amplitude of each component of $\tilde{\textbf{A}}$ is drawn from a Rayleigh distribution, which leads to a Gaussian distribution in real space, and its phase $\phi$ is randomly chosen from a uniform distribution in the interval [0,2$\pi$]. The scales $\Lambda_{\rm min,max}$ and the power-law scaling $n$ are determined empirically by comparing magnetic field simulations to the observed Faraday rotation measures of radio sources located in or behind the cluster~\citep{Coma}. We discuss our numerical choices for these parameters below. We then obtain the magnetic field in Fourier space by $\tilde{\textbf{B}}(\textbf{k})=i\, \textbf{k}\times\tilde{\textbf{A}}(\textbf{k})$ where its components are described by a power spectrum:
\begin{equation}
    |\tilde{B}_k|^2 \sim k^{-n} \, .
\end{equation}
Applying a Fourier transform we get the magnetic field in real space.
The radial profile of the magnetic field follows the gas distribution in the cluster and is added as a multiplicative factor on top of the previous randomly generated magnetic field $\textbf{B}_{\rm gen}$:
\begin{equation}
    \textbf{B} (r)=\mathcal{C} \, B_0 \left (\frac{n_e(r)}{n_{e,0}} \right )^{\eta} \textbf{B}_{\rm gen} \, ,
\end{equation}
where the exponent $\eta$ is a parameter of order-one (specific values are discussed later) which has to be fit to the data and $B_0$ is the magnetic field strength at the centre of the cluster. The radial profile of the electron density is described with a $\beta$-model~\citep{beta}:
\begin{equation}\label{edensity}
    n_e(r)=n_{e,0}\left(1+\frac{r^2}{r_c^2}\right)^{-\frac{3}{2}\beta} \, ,
\end{equation}
where $n_{e,0}$ denotes the electron density in the cluster centre, $r$ the radial distance from the centre and $r_c$ the core radius of the cluster. The normalization factor $\mathcal{C}$ ensures that the average magnetic field strength within the cluster core is equal to $B_0$ and is defined as~\citep{CAB}:
\begin{equation}\label{3D_norm}
    \mathcal{C}=\frac{N_{r<r_c}}{\sum_{r<r_c}B_{\rm gen}\cdot \frac{n_e(r)}{n_{e,0}}^{\eta}} \,  ,
\end{equation}
where $N_{r<r_c}$ is the number of lattice points within the cluster core region.

Since the Coma cluster magnetic field is the most accurately studied one, we adopt the values found for Coma $\Lambda_{\rm min}=2\, \mbox{kpc}$, $\Lambda_{\rm max}=34\, \mbox{kpc}$ and $n=11/3$ for all sources~\citep{Coma}. For $\eta$ we use the same value of 0.7 as in~\citet{non-obs} in order to compare our 3D results with the previously used 1D model. This value $\eta=0.7$ is in between the values of Coma $\eta=0.5$ and Hydra A $\eta=1.0$~\citep{hydra}. Since a larger $\eta$ means a faster drop in magnetic field strength with increasing radius, $\eta=0.7$ is a more conservative estimate than the $\eta$ derived from the Coma cluster.

In order to obtain enough samples of magnetic fields (especially for the ML methods) within a reasonable amount of time, we make use of the radial symmetry of the magnetic field and the fact that we are only interested in the field along the line of sight where the X-ray source is located. This means that we can take multiple lines of sight from each simulation. To reduce the correlation between these we require a distance of $3\, \mbox{kpc}$ (for training data) and $40\, \mbox{kpc}$ (for test data) between the lines of sight we consider. The value for the test data was chosen because of the value of $\Lambda_{\rm max}=34\, \mbox{kpc}$ which is the maximal length over which the magnetic field reverses its direction and hence, for lines of sight with a larger distance the correlation between the magnetic fields should be minimal. This was however not possible to adopt also for the training data because the computation time would be too long which is the reason for the significantly smaller distance. It is not necessary for magnetic fields used to generate training data to be completely uncorrelated since these are not used to obtain the final bounds. Furthermore, we utilize lines of sight from all three directions of the lattice. For sources which are located within the cluster we split the lines of sight in half and use both. In all cases the Pearson correlation coefficient does not show any significant correlations. 

\subsection{Comparison with the 1D model}
Instead of simulating the whole cluster magnetic field as for the 3D model, the 1D approach used in previous work emulates the magnetic field only along the line of sight to the source considered. This line of sight field is approximated by cells in which the magnetic field is constant and randomly orientated. Their lengths are drawn from a power-law distribution which is limited by $L_{\rm min}=\Lambda_{\rm min}/2$ and $L_{\rm max}=\Lambda_{\rm max}/2$. The number of cells is chosen such that it corresponds to the total propagation length of the source through the cluster. The radial profile again follows the gas distribution defined by Equation~\eqref{edensity} as for the 3D model. It must be mentioned that in contrast to the 3D model the 1D model does not generate a divergence-free magnetic field. Further, we want to emphasize that even though we are only interested in the magnetic field along the line of sight, the 3D magnetic field model is qualitatively different from the 1D model. The former does not contain discontinuities and, hence, represents a more realistic model of the real magnetic field. See~\citet{smooth1Dfields}, for an analysis of ALP conversion in smoothed out domainlike magnetic fields. An extensive analysis of the effect of different magnetic fields on ALP bounds from the AGN NGC1275 is also presented in \cite{fieldModels} (which appeared after this work).

\begin{figure*}
\centering
\begin{subfigure}{0.48\textwidth}
\includegraphics[width=\textwidth]{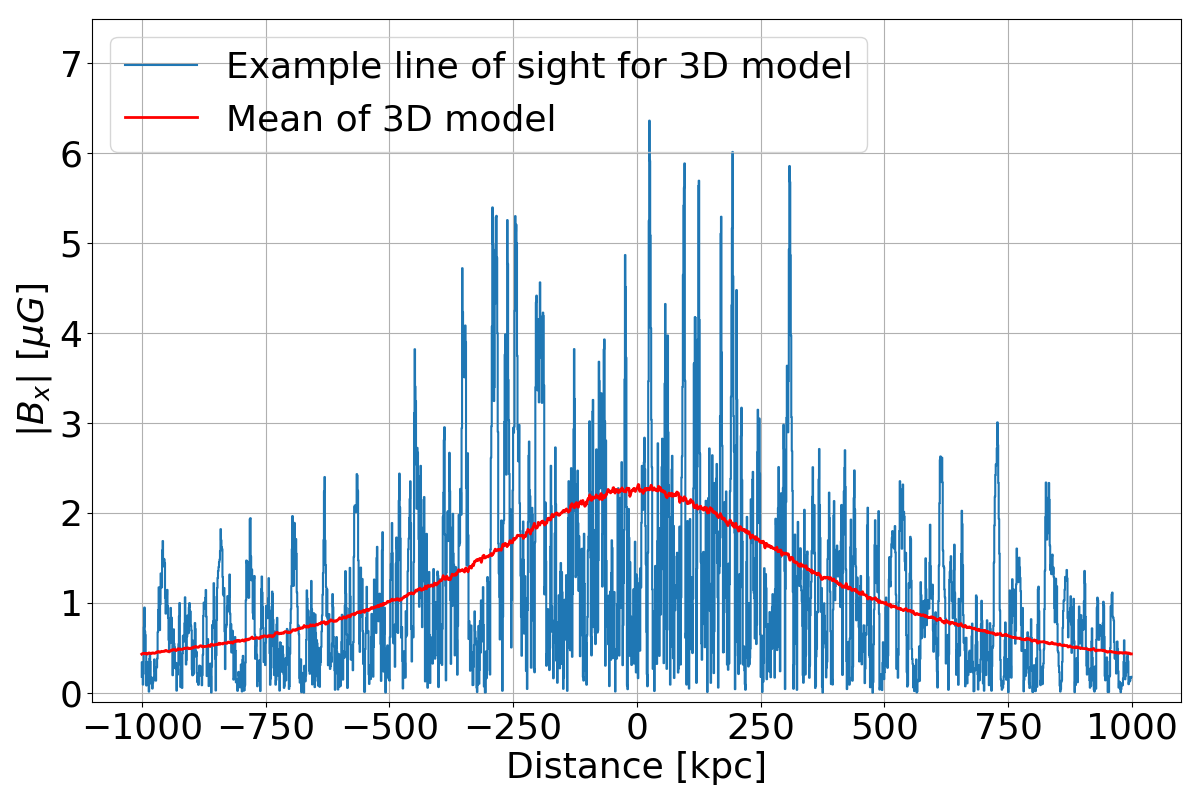}
\end{subfigure}
\begin{subfigure}{0.48\textwidth}
\includegraphics[width=\textwidth]{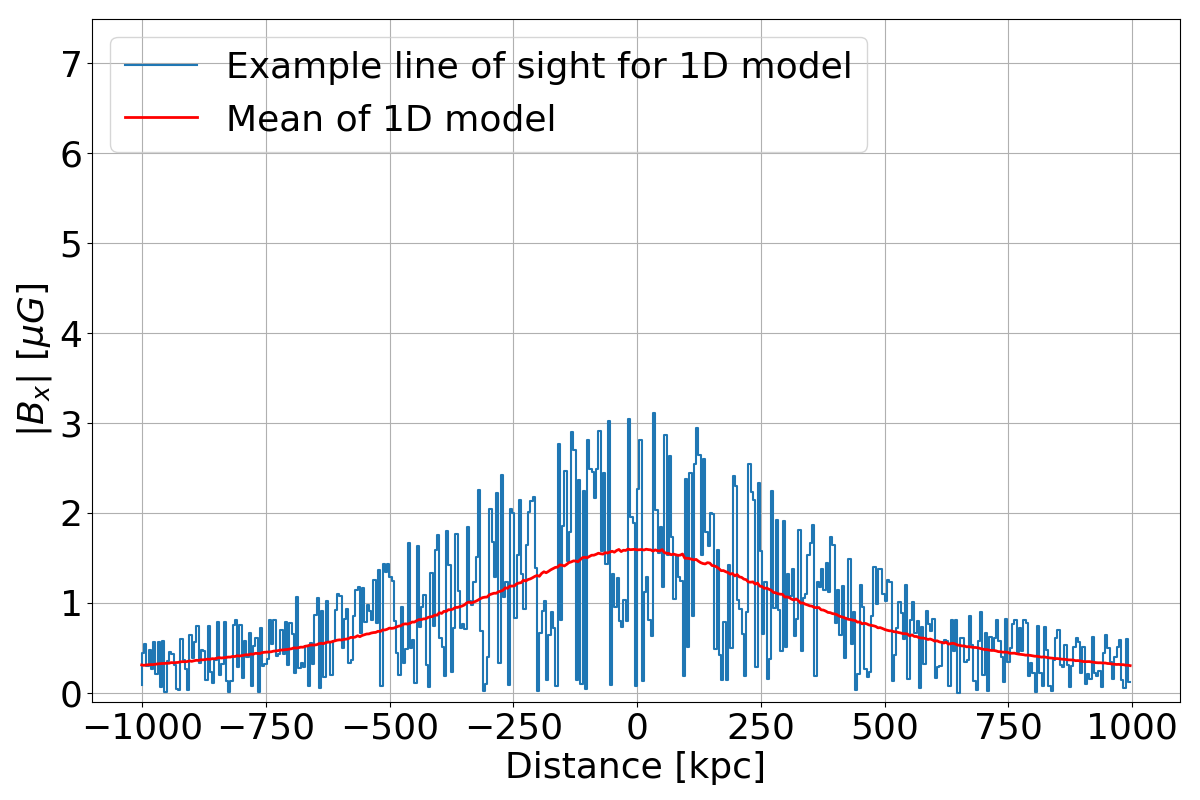}
\end{subfigure}
\begin{subfigure}{0.48\textwidth}
\includegraphics[width=\textwidth]{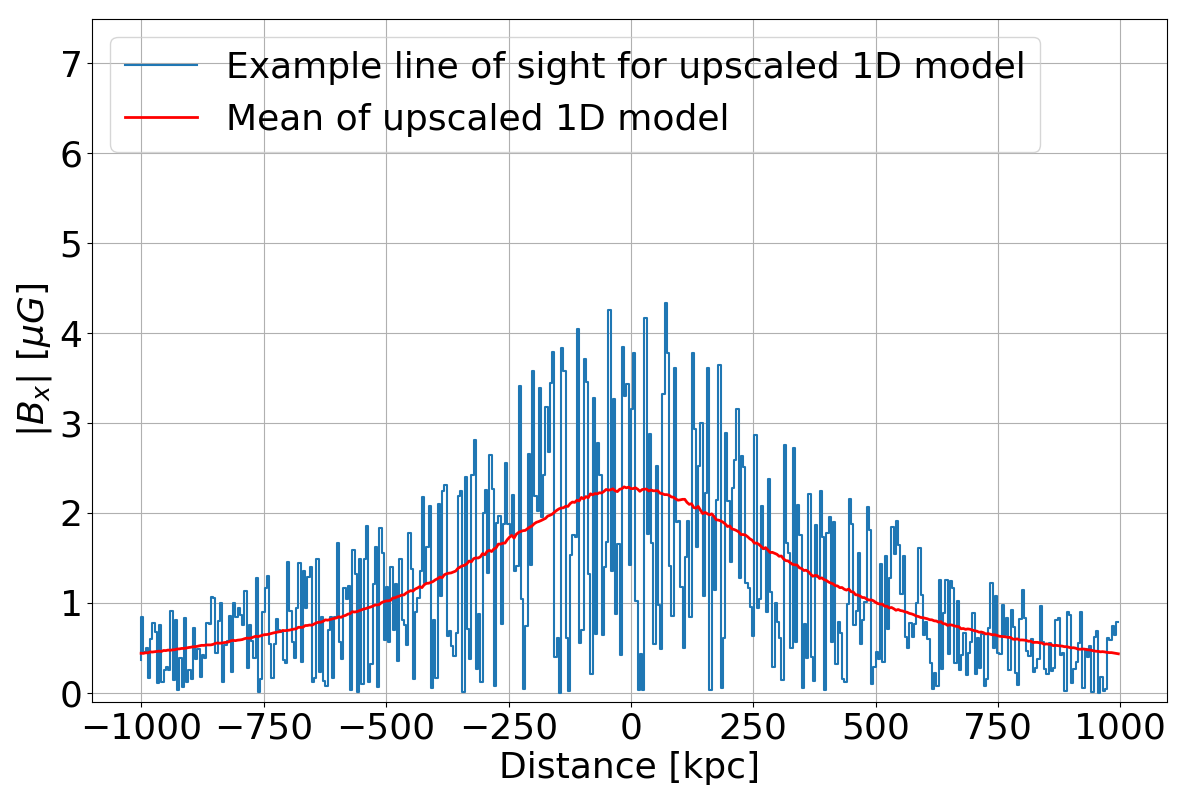}
\end{subfigure}
\begin{subfigure}{0.48\textwidth}
\includegraphics[width=\textwidth]{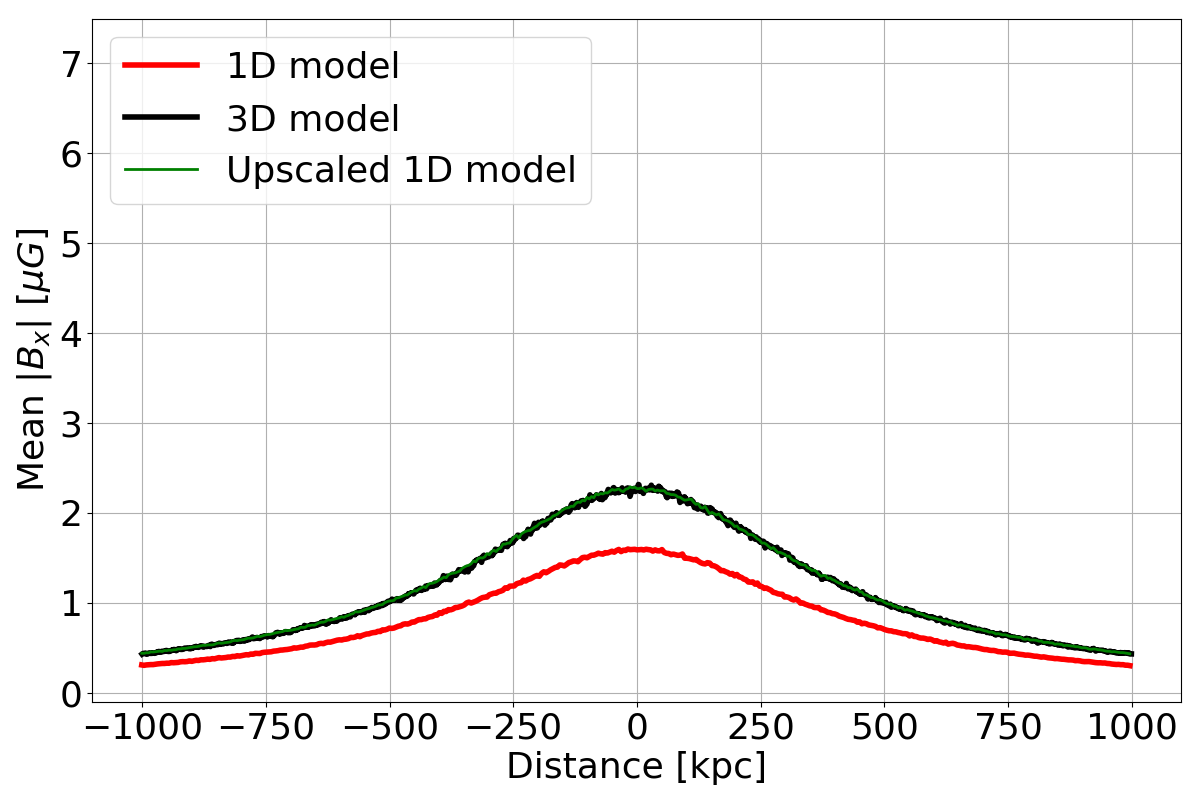}
\end{subfigure}
\caption{Examples of the magnetic fields (absolute values) along the line of sight as well as the means of all generated fields of the quasar B1256+281 behind the Coma cluster for the 3D (top left), 1D (top right) and upscaled 1D model (bottom left). The means of all models are shown in the bottom right.}
\label{fig:mag_fields}
\end{figure*}

The main difference lies within the normalization of the magnetic fields. For the 3D model we normalize it such that the mean value within the core region of the cluster equals $B_0$ as given by Equation~\eqref{3D_norm}. For the 1D model this is not possible because the magnetic field only gets simulated along the line of sight. In this case it gets normalized such that at the centre of the cluster the maximum value of the magnetic field is $B_0$. This is a more conservative estimate than for the 3D model, emerging from their different simulation methods. Therefore, we also expect the constraints from the 1D model to be more conservative than those arising from the 3D model. This can be seen in Figure~\ref{fig:mag_fields} where we have plotted examples for the magnetic fields (absolute values) along the line of sight as well as the mean of all generated magnetic fields for the quasar B1256+281 which is located behind the Coma cluster. In the two top plots this has been done for the 3D (left) and 1D model (right) where we can directly see that the different normalizations lead to larger magnetic field strengths in the 3D model.

In order to check later whether there are inherent differences between the 1D and 3D model apart from the different magnetic field strengths, we introduce an upscaled 1D model. For the latter we increase $B_0$ by hand in the 1D model in order to match the mean magnetic field of the 3D model. This is shown in the bottom left of Figure~\ref{fig:mag_fields}. However, this new model still does not fully capture all features of the 3D model. As the examples in Figure~\ref{fig:mag_fields} already indicate, the fluctuations of the 3D model are still larger than those of the upscaled 1D model. By larger fluctuations we mean that the magnetic field strength reaches larger values. In order to demonstrate this, we have calculated the mean of the maximum values of the magnetic fields as well as their mean variance across all 20,000 samples that we have generated. The results given in Table~\ref{tab:1D_3D_comp} show the discrepancy in the amplitude of the fluctuations.

\begin{table}
    \centering
    \caption{Comparison of mean maximum values and mean variance of the magnetic field strength for the 1D, 3D and upscaled 1D models.}
    \label{tab:1D_3D_comp}
    \begin{tabular}{|c||c|c|}
     \hline
    
        Model & Max $B$ [$\mu$G] & Var $B$ [$\mu$G] \\
        \hline
        \hline
        1D & 2.9 & 1.1\\
        \hline
        3D & 7.4 & 2.7\\
        \hline
        Upscaled 1D & 4.2 & 2.3 \\
        \hline
    \end{tabular}
\end{table}

\section{Astrophysical Sources}\label{sec:astro_systems}
For comparability with previous work~\citep{non-obs,Day:2019ucy}, we use the same five point sources which are either located in or behind galaxy clusters:
\begin{enumerate}
    \item The Sy1 galaxy 2E3140 within A1795 (A1795Sy1).
    \item The AGN NGC3862 within A1367 (A1367).
    \item The quasar CXOU J134905.8+263752 behind A1795 (A1795Quasar).
    \item The quasar B1256+281 behind Coma (Coma1).
    \item The quasar SDSS J130001.48+275120.6 behind Coma (Coma2).
\end{enumerate}
We use the same \textit{Chandra} ACIS observations as in~\citet{non-obs,Day:2019ucy} and the corresponding observation IDs are listed in Appendix~\ref{app:obs_ids}. In brackets we denote the abbreviations by which they are mentioned throughout this article. The observations of the sources have been processed with CIAO 4.8.1~\citep{ciao}. Multiple observations of the same source have been stacked and the background of the galaxy cluster subtracted. All these sources have already been studied using 1D magnetic field simulations with conventional statistical methods in~\citet{non-obs} and with ML methods in~\citet{Day:2019ucy} which we compare later in Section~\ref{sec:results} with our new results. 

For the first three sources we scan over a coupling range of $(0.1-2.0)\times 10^{-12} \, \mbox{GeV}^{-1}$ as in~\citet{Day:2019ucy}, whereas for the sources behind Coma we consider a range of $(1.1-3.0)\times 10^{-12} \, \mbox{GeV}^{-1}$. We use the increased range of couplings in these latter sources to be able to still obtain bounds. Even though the corresponding bounds will not be as tight as for the other sources, we still were interested in them due to the well constrained magnetic field of the Coma cluster. 

In all situations, these ranges were chosen such that for couplings at the lower end the ALP-induced oscillations are indistinguishable from the Poisson noise, whereas for couplings at the upper end the oscillations are large enough such that they would be easily detected. For A1795Sy1 and A1367 we considered an energy range of $(1-5)\, \mbox{keV}$ and for the other sources $(0.5-7)\, \mbox{keV}$. A1795Sy1, A1367 and A1795Quasar have been fitted with a power law with additional absorption from neutral hydrogen, Coma1 only with a power law and Coma2 with a power law and a Fe K$\alpha$-line ($E=6.4\, \mbox{keV}$). The parameters for the electron density and the magnetic field are the same as in~\citet{non-obs} which have been taken from~\citet{Coma,A2199,A194,eta_hydra_A,A1795B,A1795edens,A1795beta,A1367B,A1367beta}. For sources which are located behind the cluster we assume a total propagation length of $2\, \mbox{Mpc}$ whereas for sources within its host cluster we set $L_{tot}=1 \, \mbox{Mpc}$. These numbers are based on the typical size of a galaxy cluster. The exact position of the Sy1 galaxy 2E3140 within A1795 is not exactly known. We assume a midway position, but if the galaxy was actually towards the front of the cluster, the bounds would be reduced, see~\citet{non-obs} for a detailed discussion.

In Table \ref{tab:parameter_summary} we have listed all parameters that have been used for all astrophysical systems where the redshifts have been taken from SIMBAD~\citep{SIMBAD}.

\begin{table*}
\centering
\caption{Summary of the parameters for all sources.}
\label{tab:parameter_summary}
\begin{tabular}{|p{3cm}||p{2cm}|p{2cm}|p{2cm}|p{2cm}|p{2cm}|}
 \hline
 Source & 2E3140 & NGC3862 & CXOU J134905.8 +263752 & B1256+281 & SDSS J130001.47 +275120.6 \\
 \hline
 Cluster & A1795 & A1367 & A1795 & Coma & Coma \\
 \hline
  z$_{source}$ & 0.059  & 0.0216 & 1.30 & 0.38 & 0.975 \\
 \hline
  z$_{cluster}$ & 0.063 & 0.0225 & 0.063 & 0.023 & 0.023 \\
 \hline
  Offset (kpc) & 456 & 186 & 194 & 232 & 215 \\
 \hline
  $L_{tot}$ (Mpc) & 1 & 1 & 2 & 2 & 2 \\
 \hline
  $\Lambda_{\rm min}$ (kpc) & 2 & 2 & 2 & 2 & 2 \\
 \hline
  $\Lambda_{\rm max}$ (kpc) & 34 & 34 & 34 & 34 & 34 \\
 \hline
  $\eta$ & 0.7 & 0.7 & 0.7 & 0.7 & 0.7 \\
 \hline
  $n$ & 11/3 & 11/3 & 11/3 & 11/3 & 11/3 \\
 \hline
  $B_0$ ($\mu$G) & 20 & 3.25 & 20 & 4.7 & 4.7 \\
 \hline
  $n_{e,0}$ (10$^{-3}$ cm$^{-3}$) & 50 & 1.15 & 50 & 3.44 & 3.44 \\
 \hline
  $r_c$ (kpc) & 146 & 308 & 146 & 291 & 291 \\
 \hline
  $\beta$ & 0.631 & 0.52 & 0.631 & 0.75 & 0.75 \\
 \hline
\end{tabular}
\end{table*}

\section{Methods}\label{sec:methods}
Below we describe the four different methods used in this paper to constrain the ALP-photon coupling. The basic idea of all methods is to compare the response of the real spectrum as observed by \textit{Chandra} to fake spectra with or without ALPs. These fake spectra can be produced with \textit{Chandra's} software package \textit{Sherpa}. \textit{Sherpa}'s function \textit{fake\_pha} generates spectra with a certain source model and random Poisson noise that is determined by the exposure time which we set equal to the observation time of the real spectrum. \textit{fake\_pha} also simulates instrumental effects such as the detector's finite energy resolution ($\sim 150$ eV). As source model for spectra without ALPs we use the function with which the real spectra have been fitted:
\begin{equation}
    F_0(E)=AE^{-\gamma} \left(\times~{\rm e}^{-n_H\sigma(E(1+z))}\right) \left(+~{\rm Fe} \, {\rm K}\alpha\right) \, .
\end{equation}
The first part corresponds to the power law model (free parameters are the amplitude $A$ as well as the exponent $\gamma$) with which every source is fitted. The second part denotes the absorption from neutral hydrogen (free parameter column density $n_H$), whereas the last part indicates the Fe K$\alpha$ line (given in brackets since these are not applied for all sources, see Section~\ref{sec:astro_systems} for details). For spectra with ALPs we multiply model $F_0$ with the simulated photon survival probability:
\begin{equation}
    F_1(E,\textbf{B},g_{a\gamma\gamma})= F_0(E) \times P_{\gamma \rightarrow \gamma}(E(1+z),\textbf{B},g_{a\gamma\gamma}) \, .
\end{equation}

\subsection{Bounds from a \texorpdfstring{$\chi^2$}{} statistic}
We follow the exact procedure from~\citet{non-obs}. We start by generating 1,000 fake spectra for each coupling. We then fit each spectrum with the model $F_0$ and compute the reduced $\chi^2$-statistic. Finally, we compare it with the $\chi^2_{red}$ of the real spectrum and if 95\% of the fake spectra lead to a worse fit (i.e.~a larger $\chi^2_{red}$) than the real data, we can exclude $g_{a\gamma\gamma}$ at a 95\% confidence level. For sources where $\chi^2_{red}$ is smaller than 1, we only consider spectra with $\chi^2_{red}>1$ as a worse fit.
\subsection{Single coupling ML method}
This method refers to the procedure presented in~\citet{Day:2019ucy}. Its main idea is to train a ML classifier to distinguish between fake spectra without ALPs and spectra with ALPs of a certain coupling strength. For all upcoming three ML methods we use training sets with 8,000 and test sets with 2,000 samples. In order to generate these we have simulated the photon survival probability for each source 800 and 200 times respectively where we have always used a different magnetic field configuration. From each survival probability we have generated 10 fake spectra which differ in their Poisson noise.

Since the photon survival probability is always smaller or equal to 1, the fake spectra with ALPs have an overall lower flux than spectra without ALPs, a difference that would be readily picked up by the classifiers. Hence, we cannot take these spectra as input data. This is because when fitting AGN spectra, the overall amplitude of the spectrum is a free parameter, so cannot be used to test for the presence of ALPs. Therefore, we use three different data products. First, we fit every fake spectrum with the model $F_0$ and use the residuals (resid). Second, we generate fake spectra with source model $F_1$ multiplied with the inverse, average survival probability (up). This results in upscaled spectra which do not suffer from a flux reduction due to the ALP-photon interconversion but still contain the oscillatory features. Third, we can refit the upscaled spectra with $F_0$ and store the residuals (up-resid). In brackets we have denoted the abbreviations by which the different data products will be referred to in the upcoming sections.

We use various different classifiers from \textit{scikit-learn}~\citep{sklearn} (version 0.23.2): Quadratic Discriminant Analysis (QDA), Gaussian Naive Bayes (GNB), Decision Tree Classifier (DTC), Random Forest Classifier (RFC), AdaBoost Classifier (ABC) and Support Vector Machine Classifier (SVM). We use a grid search in order to optimize their hyperparameters. This is performed for one specific coupling and we adopt the values for all other couplings. Furthermore, we perform no additional grid search for the up-resid data product since this data is quite similar to the resid type. More details are listed in Appendix~\ref{app:hyperparameters}. In order to use these classifiers to constrain ALPs, we have to define a test statistic for a data set $\mathcal{D}$:
\begin{center}
    TS$_{\mathcal{D}}$ = highest value of $g_{a\gamma\gamma}$ such that $\mathcal{C}_g$ classifies $\mathcal{D}$ as ALPs \, ,
\end{center}
where $\mathcal{C}_g$ is a classifier trained on a specific coupling $g$.
Furthermore, we have to define a null hypothesis:
\begin{center}
    $H_0$ = ALPs exist with $g_{a\gamma\gamma}=g_{\rm null}$ \, .
\end{center}
We can then feed all test sets to all the classifiers and thereby obtain the null distribution. If 95\% of TS$_{\mathcal{D}^i(g_{\rm null})}$ are larger than TS$_{\rm real}$, the test statistic for the real data, we can exclude $g_{\rm null}$ at a 95\% confidence level.

\begin{figure*}
\centering
\begin{subfigure}{0.48\textwidth}
\includegraphics[width=\textwidth]{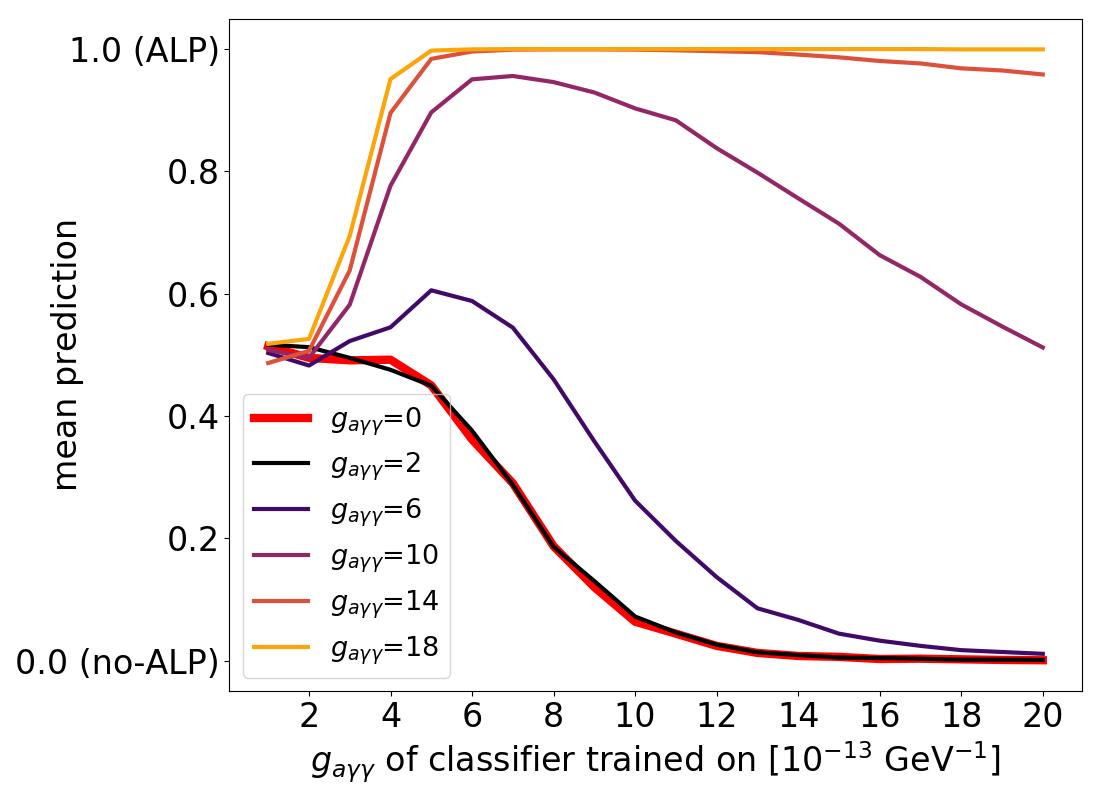}
\end{subfigure}
\begin{subfigure}{0.48\textwidth}
\includegraphics[width=\textwidth]{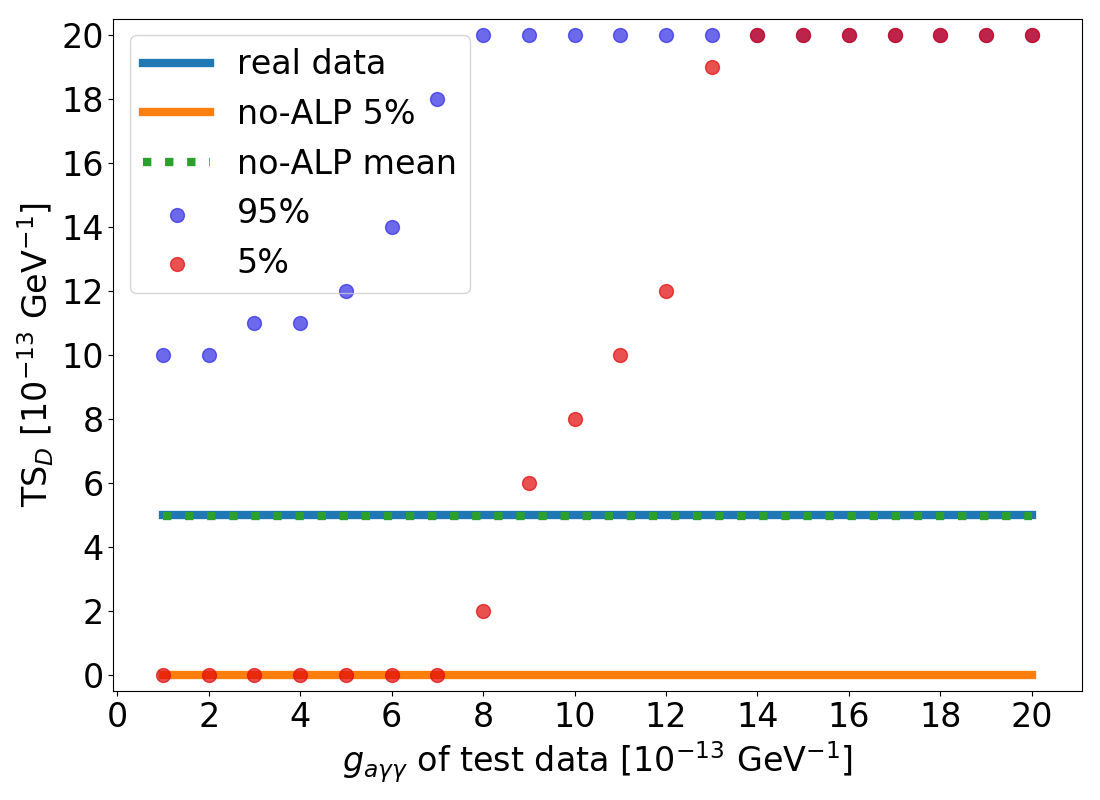}
\end{subfigure}
\caption{\textit{Left:} Performance of the QDA classifier trained on the resid data of A1795Sy1. The colors denote different couplings whose test sets are fed to the different classifiers. The couplings $g_{a\gamma\gamma}$ in the legend are given in $10^{-13} \, \mbox{GeV}^{-1}$. \textit{Right:} Test statistic quantiles for the different test sets, as well as the test statistic of the real spectrum for the same classifier and data as on the left.}
\label{fig:single_ml}
\end{figure*}

To illustrate the performance of our classifiers, we plot the performance of the QDA classifier trained on the resid data of A1795Sy1 on the left of Figure~\ref{fig:single_ml}. The $x$-axis shows the coupling on which the classifier is trained and the $y$-axis denotes the mean prediction where 0 refers to no ALPs and 1 to ALPs. Classifiers which are trained on small couplings ($g_{a\gamma\gamma}\leq 2 \times 10^{-13} \, \mbox{GeV}^{-1}$) return a mean prediction of 0.5 for all test sets. Hence, in this coupling regime the no-ALP and ALP data show no differences, i.e.~the Poisson noise is larger than the ALP-induced oscillations. This can be also seen by looking at the curves of the no-ALP data ($g_{a\gamma\gamma} = 0$) and the ALP data with $g_{a\gamma\gamma} = 2 \times 10^{-13} \, \mbox{GeV}^{-1}$ which match almost perfectly. Their mean prediction approaches $0$ as the coupling on which the classifiers have been trained increases. Test sets with large couplings ($g_{a\gamma\gamma} \geq 14 \times 10^{-13} \, \mbox{GeV}^{-1}$) are very well classified as ALP data if the coupling on which the classifiers have been trained is larger than $5 \times 10^{-13} \, \mbox{GeV}^{-1}$. Test sets of intermediate couplings ($3 \times 10^{-13} \, \mbox{GeV}^{-1} \leq g_{a\gamma\gamma} \leq 13 \times 10^{-13} \, \mbox{GeV}^{-1}$) have a maximum which is located at a coupling equal or slightly smaller than its own. After the maximum the mean prediction drops because for classifiers trained on larger couplings the smaller oscillations of the test sets with intermediate couplings are not as large as those on which they have been trained on and therefore are not as easy to detect.

The right of Figure~\ref{fig:single_ml} shows the 5th and 95th percentile of the test statistic of the ALP test sets for the QDA classifier trained on the resid data of A1795Sy1. Additionally, we plot the 5th percentile and the mean of the no-ALP test set as well as the test statistic of the real data. The constraint on $g_{a\gamma\gamma}$ (at a 95\% confidence level) corresponds to the value on the $x$-axis where the 5-percentiles of the ALP test sets cross the line of the real spectrum.

\subsection{Approximate Bayesian computation}
\begin{figure}
\centering
\includegraphics[width=\columnwidth]{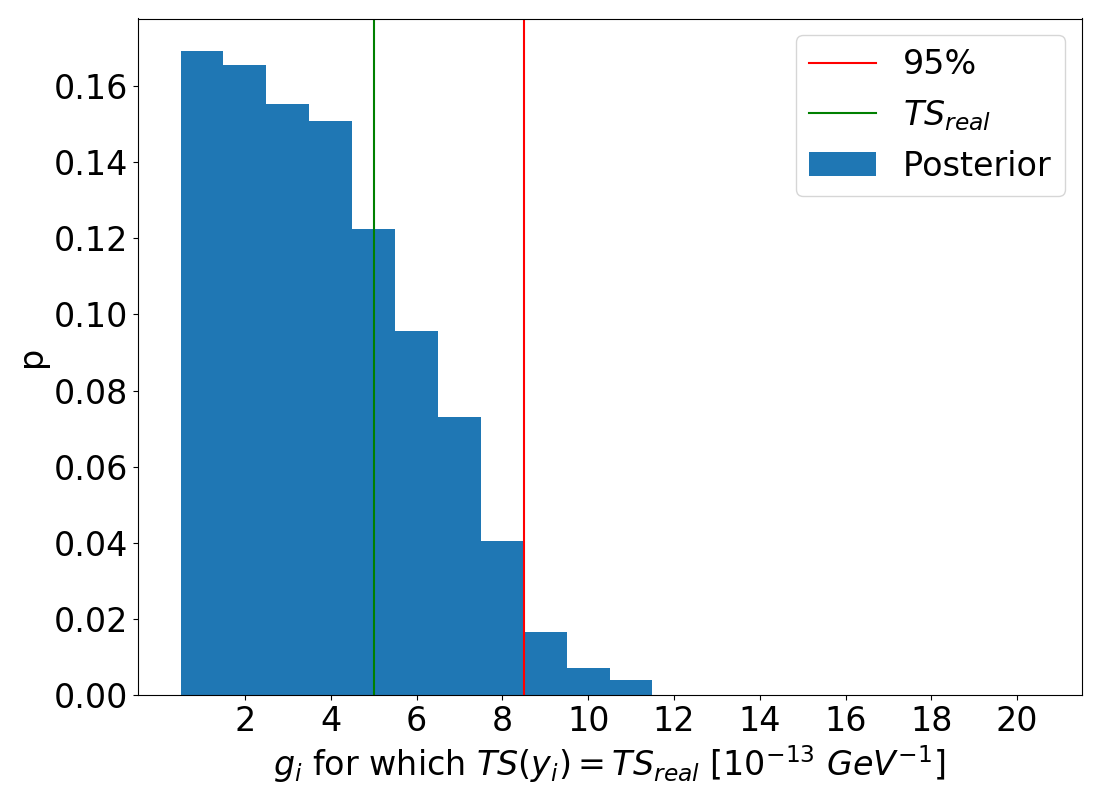}
\caption{ApBC-approximated posterior distribution and its 95th percentile for the QDA classifiers of the Sy1 galaxy 2E3140 within A1795.}
\label{fig:ABC}
\end{figure}

ApBC is an inference method used when the likelihood either cannot be calculated or would be computationally too expensive. Therefore, it has to be simulated based on the prior probability distribution. For more details and a good overview see~\citet{abc}. In our case, we simply assume a uniform prior across the couplings $g_{a\gamma\gamma}$ considered (cf.~Section~\ref{sec:astro_systems}) and use the classifiers and the test statistic from the previous method. We then perform the following three steps:
\begin{enumerate}
    \item Feed all test sets into all classifiers.
    \item Calculate the test statistic.
    \item If the test statistic of the test data is the same as for the real data, we accept the coupling $g_{a\gamma\gamma}$ of the test data.
\end{enumerate}
As output we obtain a set of $g_{a\gamma\gamma}$ sampled from the posterior distribution $\pi(g_{a\gamma\gamma}|y_{{\rm real}})$ which can be used to approximate the posterior. Values for $g_{a\gamma\gamma}$ larger than the 95th percentile of the approximated posterior distribution can then be excluded at a 95\% confidence level. In Figure~\ref{fig:ABC} we plot an example of an ApBC-approximated posterior distribution and its 95th percentile as well as the test statistic of the real data for the QDA classifiers trained on the resid data of A1795Sy1.

\subsection{Multiclass classification method}

\begin{figure*}
\centering
\begin{subfigure}{0.48\textwidth}
\includegraphics[width=\textwidth]{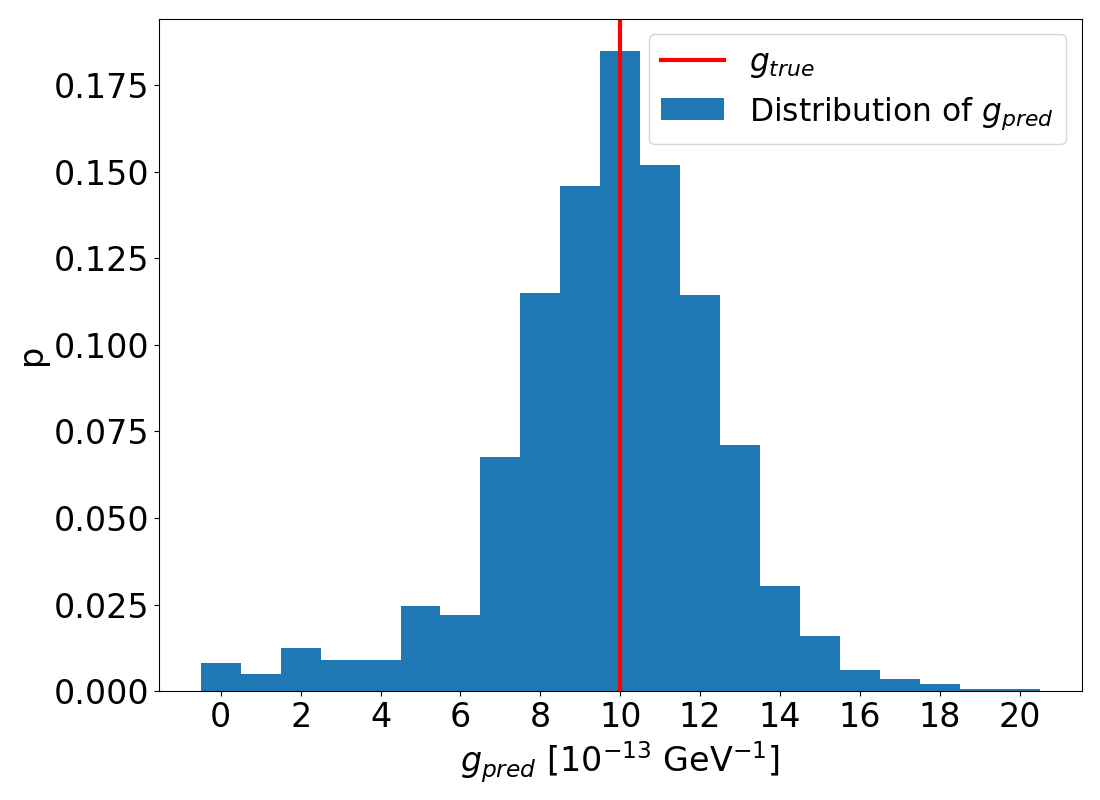}
\end{subfigure}
\begin{subfigure}{0.48\textwidth}
\includegraphics[width=\textwidth]{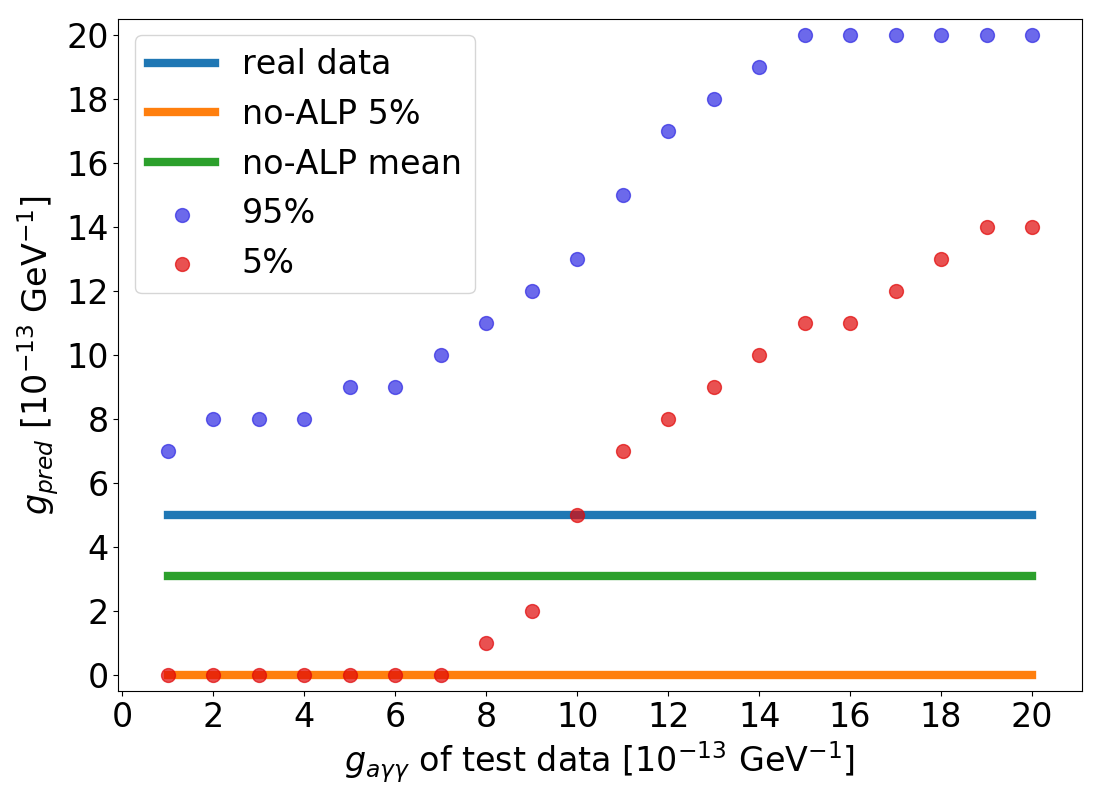}
\end{subfigure}
\caption{\textit{Left:} The distribution of predicted values of the resid data for A1795Sy1 where the true coupling constant is $g_{a\gamma\gamma}= 10^{-12} \, \mbox{GeV}^{-1}$. The QDA classifier is used. \textit{Right:} The 5th and 95th percentiles of the distribution of $g_{\rm pred}$ as well as $g_{\rm pred,real}$ and the predicted couplings of the no-ALP data (mean and 5th percentile).}
\label{fig:multi_ml}
\end{figure*}

Multiclass classification has been already suggested in~\citet{stat-impr} to constrain ALPs. Instead of training a new classifier for each coupling to distinguish between ALP and no-ALP data we build one classifier which tries to predict the exact coupling of each data set. Since the differences of two data sets with similar couplings are very subtle the performance of the classifier will be far from perfect. However, we hope that the distribution of predicted couplings will be distributed around the true coupling. As an example we plot the distribution of predicted values of the resid data for A1795Sy1 (QDA) on the left in Figure~\ref{fig:multi_ml}. From that we can see that the predicted couplings indeed are distributed around the true value. We can then use this multiclass classifier to place bounds on ALPs by applying the following procedure where we use the predicted couplings as a test statistic: 
\begin{enumerate}
    \item Predict the coupling of the real data $g_{\rm pred,real}$.
    \item Predict the couplings for the test data of all couplings $g_{{\rm pred},D^i}$.
    \item If 95\% of the $g_{{\rm pred},D^i}$ are larger than $g_{\rm pred,real}$, the coupling of the corresponding test data is excluded at a 95\% confidence level.
\end{enumerate}

The right of Figure~\ref{fig:multi_ml} shows these results for the same example as the plot on the left. As for the single coupling ML method the bound on $g_{a\gamma\gamma}$ (at a 95\% C.L.) is represented by the value on the $x$-axis where the red dots cross the blue line.\\
With this method we use the Quadratic Discriminant Analysis (QDA) from \textit{scikit-learn} as well as a deep neural network (DNN) implemented within \textit{Keras}~\citep{keras}. The architecture of our neural network is shown in Table~\ref{tab:dnn}.

\begin{table*}
\centering
\caption{Architecture of the deep neural network that we apply for the multiclass classification method. We use categorical cross-entropy as loss function, a batch size of 32 and the Adam optimizer with a learning rate of 0.0001.}
\label{tab:dnn}
\begin{tabular}{c c c c}
 \hline
 Type of Layer & Dimension & Activation & Initializer \\
 \hline
 \multirow{2}{*}{Input} & Number of energy bins & & \\
 & (source dependent) & &  \\
 Dense & 80 & SELU & \texttt{lecun\_initializer} \\
 Dense & 70 & SELU & \texttt{lecun\_initializer} \\
 Dense & 60 & SELU & \texttt{lecun\_initializer} \\
 Dense & 50 & SELU & \texttt{lecun\_initializer} \\
 Dense & 40 & SELU & \texttt{lecun\_initializer} \\
 Dense & 21 & softmax & \\
 \hline
\end{tabular}
\end{table*}

\section{Results}\label{sec:results}
We apply the methods presented in the previous section to the five different sources presented in Section~\ref{sec:astro_systems}. We have listed the constraints on $g_{a\gamma\gamma}$ in \cref{tab:bounds_A1795Sy1,tab:bounds_A1367,tab:bounds_A1795Quasar,tab:bounds_Coma1,tab:bounds_Coma2} which can be found at the end of this paper. All reported constraints are at a 95\% confidence level unless otherwise stated. Here, we discuss the results from different perspectives:

\begin{enumerate}

\item {\bf 1D vs.~3D magnetic field model:} In order to compare the bounds arising from a 1D and 3D magnetic field model, we use the bounds from the $\chi^2$ statistic. In Table~\ref{tab:bounds_1D_3D_comp} we list the constraints from all sources where those for the 1D model are taken from~\citet{non-obs}. As we can see the 3D model returns tighter bounds throughout all sources. This improvement is not surprising as we have seen in Section~\ref{sec:mag_fields} that the 1D model uses a different normalization of the magnetic field which leads to an overall weaker field. In order to check whether the better bounds arise from the differences in the strength of the magnetic field, we derive the constraint using the upscaled 1D model presented in Section~\ref{sec:mag_fields} for the source A1795Sy1. For that we obtain $g_{a\gamma\gamma} \lesssim 1.1 \times 10^{-12} \, \mbox{GeV}^{-1}$ which is already closer to the constraint from the 3D model but still not as tight. In Section~\ref{sec:mag_fields} we have seen that even though the mean of the upscaled 1D and the 3D model match perfectly the amplitude in the magnetic field strength of the 3D model is still larger. Therefore, we argue that probably these larger amplitudes are responsible for the tighter constraints.

\begin{table}
\centering
\caption{Constraints on the ALP-photon coupling in units of $10^{-12}\, \mbox{GeV}^{-1}$ for a 1D and 3D magnetic field model from the $\chi^2$ statistic.}
\label{tab:bounds_1D_3D_comp}
\begin{tabular}{|c|| c| c|}
\hline
 Sources & 1D model & 3D model \\
\hline
\hline
A1795Sy1 & 1.5 & 0.9 \\
\hline
A1367 & 2.4 & 2.0 \\
\hline 
A1795Quasar & 10.0 (75\% C.L.) & 1.3 (87\% C.L.) \\
\hline
Coma1 & 6.0 & 2.5 \\
\hline
Coma2 & 10.0 (87\% C.L.) & 3.0 (90\% C.L.) \\
 \hline
\end{tabular}
\end{table}

\item {\bf Source comparison:} The bounds from the different sources depend on two things: The quality of the spectrum and the amplitude of the potential ALP-induced oscillations. The former relies on the observation time, the redshift and the source luminosity. The larger the observation time and luminosity and the smaller the redshift the better the quality of the spectrum. Thus, the Poisson noise becomes less which makes the ALP-induced oscillations easier to detect.

From Equation~\ref{eom} we know that the amplitude of the oscillations depends on the magnetic field which relies on its strength in the center of the galaxy cluster $B_0$\footnote{Obviously, the magnetic field depends on more parameters but apart from the central magnetic field strength and the electron density, those parameters are assumed to be equal for all sources.}, and the cluster electron density $n_{e}$. The larger the central magnetic field, the larger the ALP-induced oscillations. On the other hand, a larger electron density suppresses the interconversion of ALPs and photons. From Equation~\ref{edensity} we can see that the electron density depends on three parameters: A smaller $n_{e,0}$, a smaller $\beta$ and a larger $r_c$ lead to a smaller electron density and thus, to larger ALP-induced oscillations.

Another influence is the position of the source: If it lies behind the galaxy cluster it has twice the propagation length in comparison to sources which are located within the cluster. Furthermore, the distance of the source with respect to the cluster centre impacts the size of the oscillations. A source with a position further away from the centre experiences a smaller magnetic field and therefore, smaller ALP-induced oscillations.

Out of the five sources, the two within/behind A1795 gave the best constraints (at least for the ML methods). The main reason for that is the strong magnetic field of A1795 which is larger by a factor of 6 compared to that of A1367 and 4 compared to the Coma cluster.

\item {\bf Data product comparison:} 
For the source A1795Sy1 the constraints are homogeneous across different data products. For A1367 the bounds from the up-resid and up data are significantly better than for the resid data. For all quasars behind A1795 and Coma we do not obtain any constraints for the up data. This happens due to the large test statistic of the real data, i.e.~all classifiers trained on the up data of these three sources classified the real spectrum as maximally `axiony'. In the case of the multiclass method, the real data gets classified as data with the largest possible coupling.

\item {\bf Classifier comparison:} 
In order to compare the classifiers, we count how many times the respective classifier provides the best constraint across different data products and sources using the single coupling ML method: 1. DTC (5$\times$); 2. SVM (3$\times$); 3. RFC (2$\times$); 4. ABC, QDA (1$\times$). For the approximate Bayesian computation method we find: 1. QDA, RFC, ABC (4$\times$); 4. SVM (3$\times$); 5. GNB (2$\times$); 6. DTC (1$\times$).

Interestingly, for the single coupling ML method the DTC gives more often better constraints than the RFC or the ABC. This is surprising since the RFC and ABC are improved algorithms which are based on the DTC and hence, should perform better.

In order to understand why this happens we show in Figure~\ref{fig:dtc_rfc_comp} the comparison of the performance and bounds plots for the DTC and the RFC trained on the resid data of A1795Sy1. From the two performance plots we can see that the RFC actually performs slightly better than the DTC. This leads to a different behaviour of the five-percentiles of the test statistic for the ALP data: For g$_{a\gamma\gamma} > 1.3 \times 10^{-12} \, \mbox{GeV}^{-1}$ of the test data, the five-percentiles of the RFC are larger than the corresponding ones of the DTC since the former predicts the test set to be more `axiony' (mean predictions are larger) than the DTC. For smaller couplings however, the five-percentiles of the test statistic of the DTC are larger than the corresponding ones of the RFC because for those couplings the DTC is not as good as the RFC in classifying them as no-ALP data. Theoretically, the real data should then also have a larger test statistic for the DTC when we assume that it follows the no-ALP data, i.e.~one would expect that it should be close to the mean of the no-ALP data (plotted as the green line in the two middle plots of Figure~\ref{fig:dtc_rfc_comp}). However, we can see that it is much lower than that. Therefore, the DTC delivers a better bound on ALPs because the five-percentiles of the test data cross the test statistic of the real data at a much smaller coupling of the test data. Admittedly, this is problematic because it only comes up due to the weaker performance of the DTC on low couplings. This is also the case for the other sources or data products where the DTC gives surprisingly good constraints. Fortunately, the ApBC method is able to circumvent this problem. The two bottom plots in Figure~\ref{fig:dtc_rfc_comp} show the ApBC distributions of the DTC and RFC. The difference is that for the DTC more test sets of intermediate couplings have the same test statistic as the real data than for the RFC. This happens due to the worse performance of the DTC on test sets with those couplings. As an example we consider the curve for $g_{a\gamma\gamma}=10 \times 10^{-13} \, \mbox{GeV}^{-1}$ in the two top plots of Figure~\ref{fig:dtc_rfc_comp}. For the DTC this curve is below that of the RFC. Hence, the probability that the test statistic is equal to $4\times 10^{-13} \, \mbox{GeV}^{-1}$ (the test statistic of the real data for the DTC) is larger. This leads to more samples with higher couplings that have the same small test statistic as the real data and in the end to worse constraints on ALPs for the DTC than the RFC.

\begin{figure*}
\centering
\begin{subfigure}{0.48\textwidth}
\includegraphics[width=\textwidth]{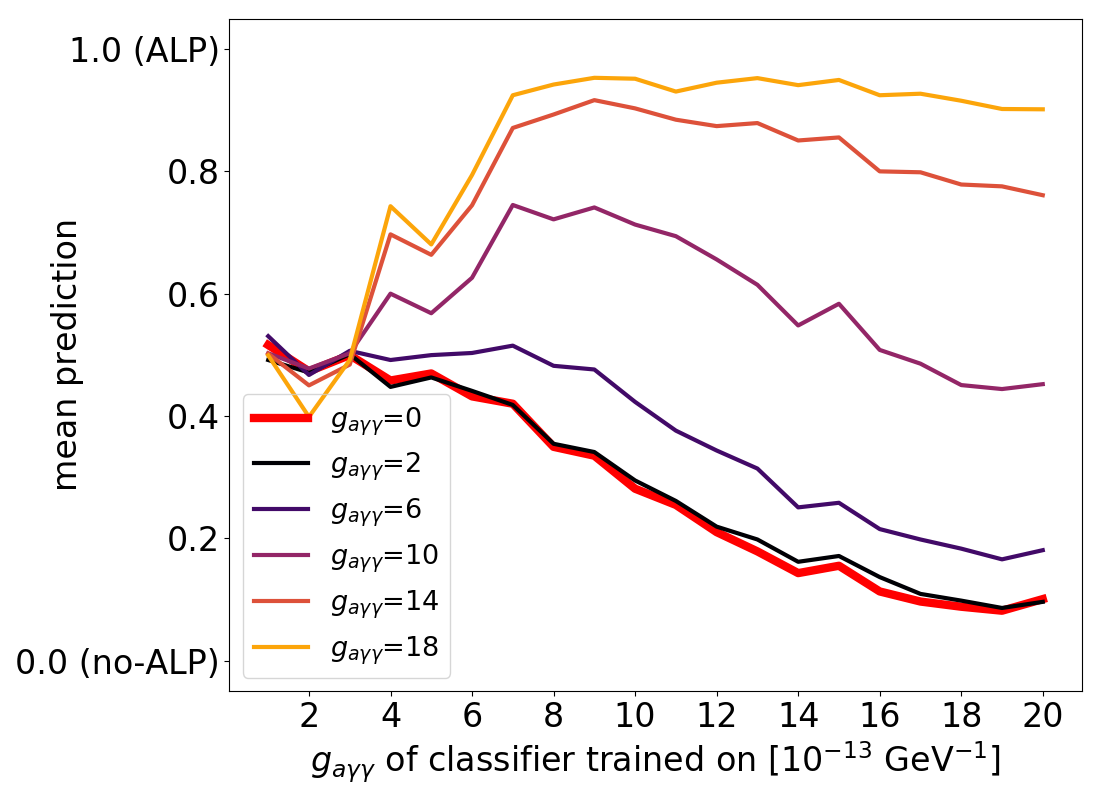}
\end{subfigure}
\begin{subfigure}{0.48\textwidth}
\includegraphics[width=\textwidth]{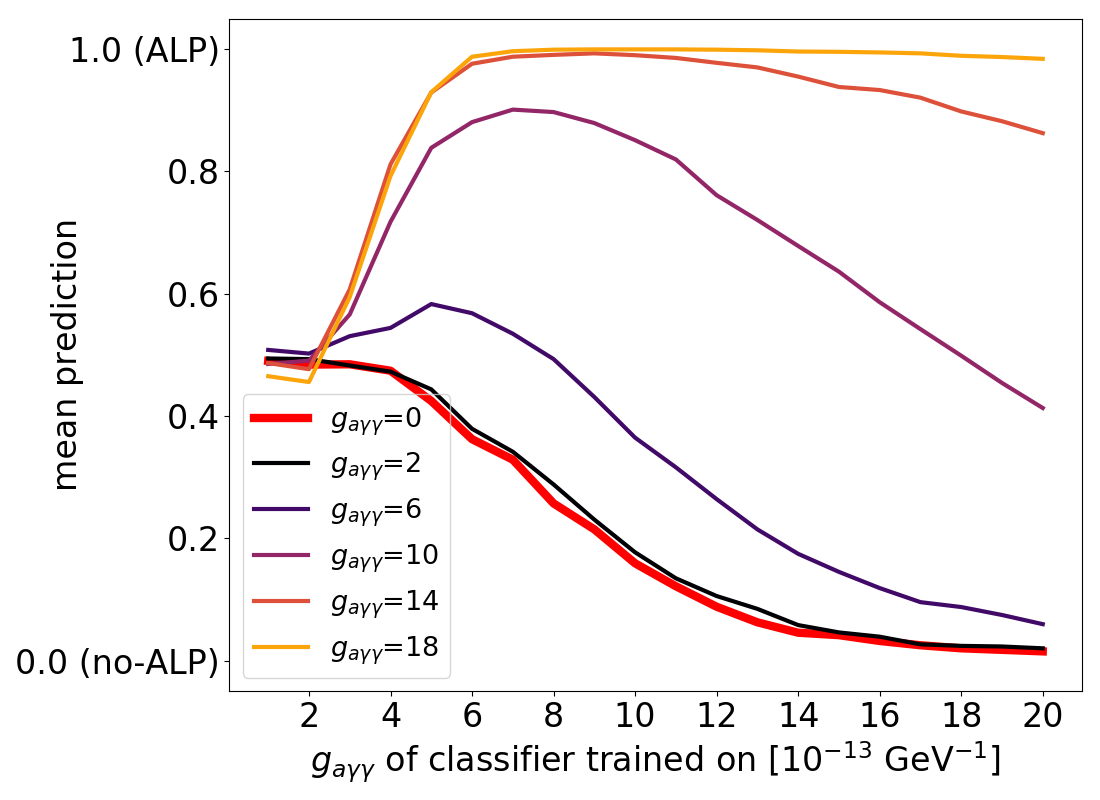}
\end{subfigure}
\begin{subfigure}{0.48\textwidth}
\includegraphics[width=\textwidth]{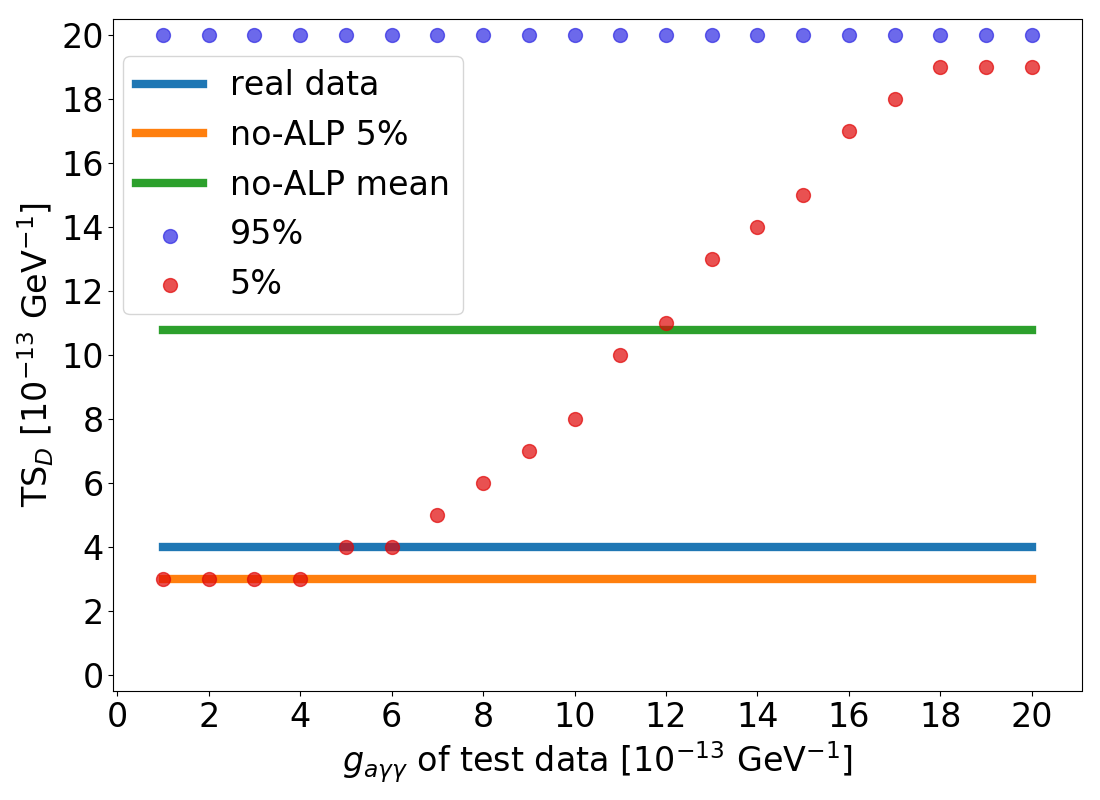}
\end{subfigure}
\begin{subfigure}{0.48\textwidth}
\includegraphics[width=\textwidth]{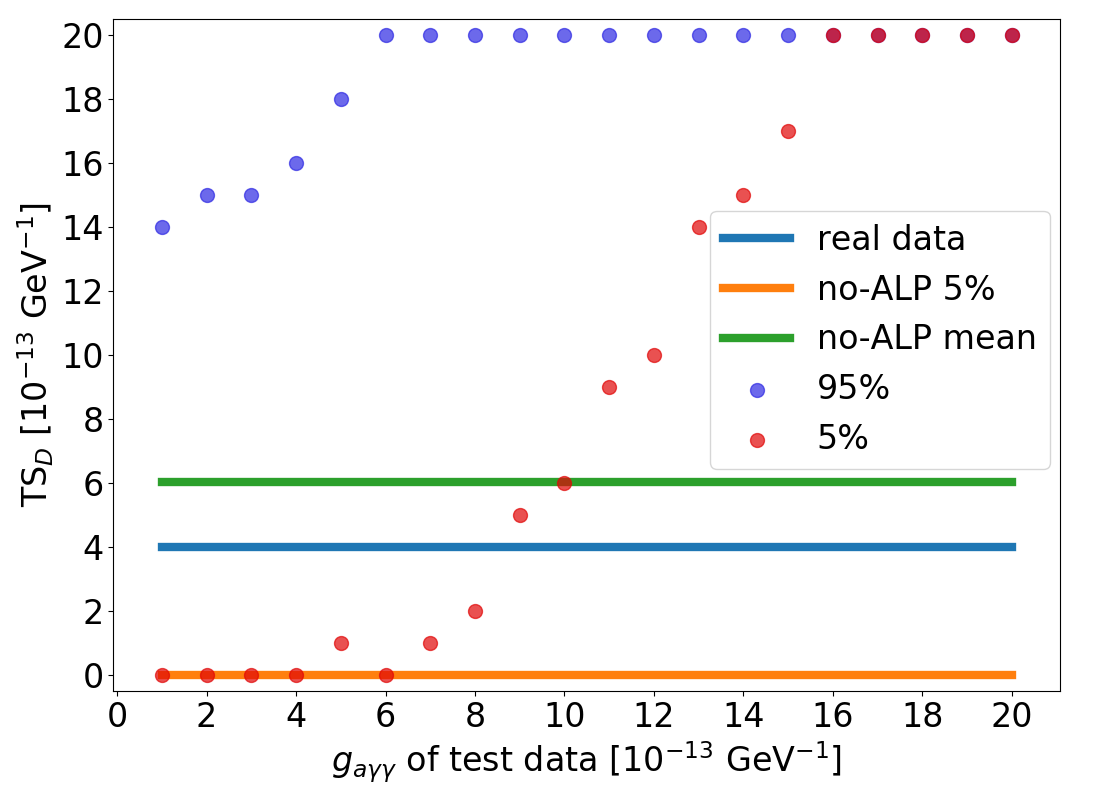}
\end{subfigure}
\begin{subfigure}{0.48\textwidth}
\includegraphics[width=\textwidth]{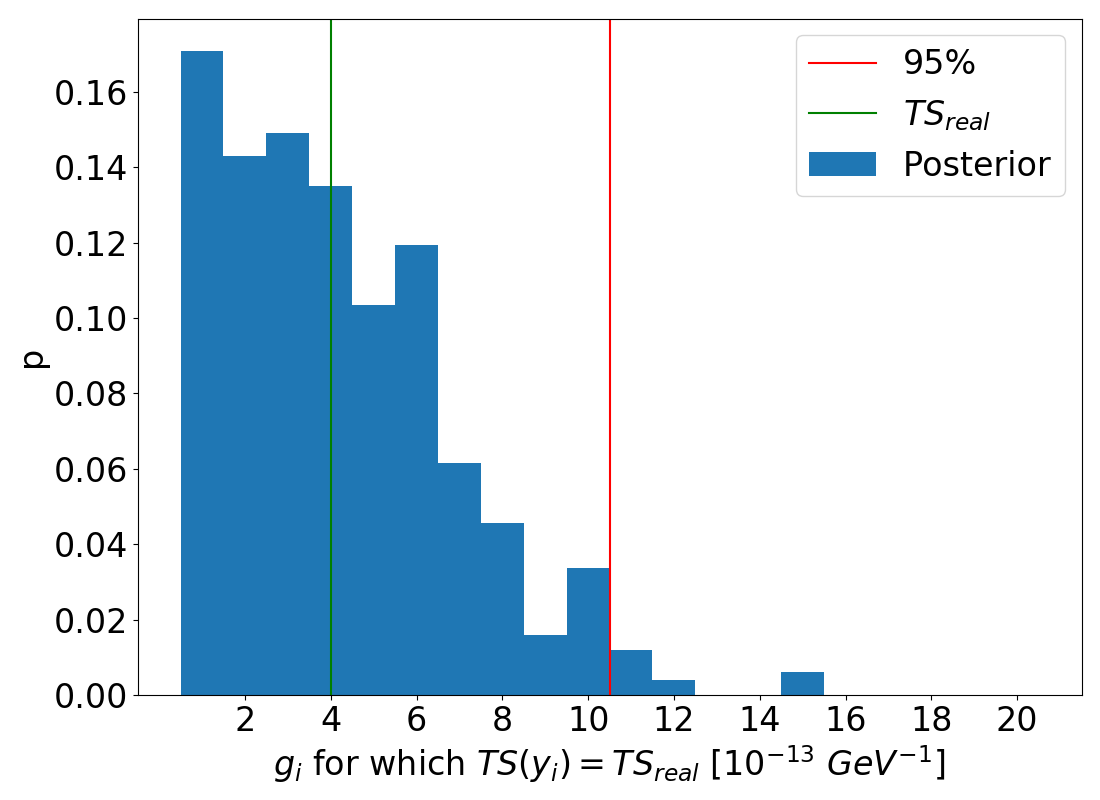}
\end{subfigure}
\begin{subfigure}{0.48\textwidth}
\includegraphics[width=\textwidth]{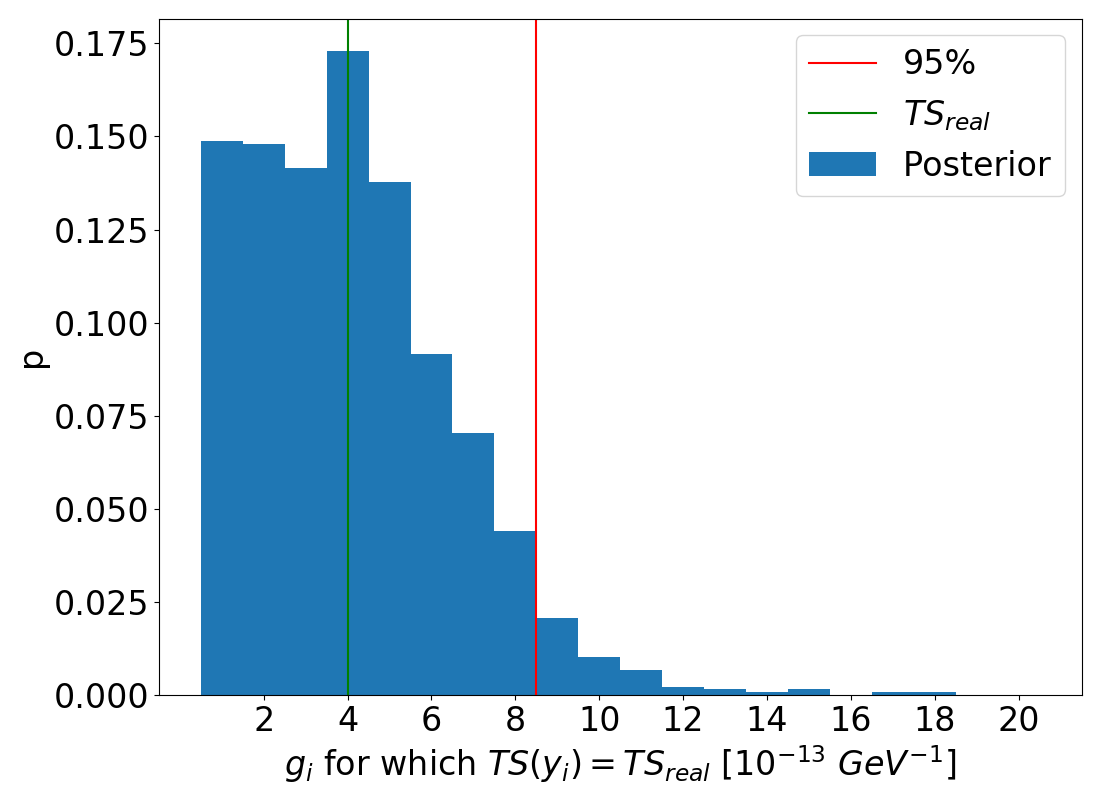}
\end{subfigure}
\caption{\textit{Top left:} Performance of the DTC. \textit{Top right:} Performance of the RFC. \textit{Centre left:} Bounds plot of the DTC. For the plots in the top, the couplings $g_{a\gamma\gamma}$ in the legend are given in $10^{-13} \, \mbox{GeV}^{-1}$. \textit{Centre right:} Bounds plot of the RFC. \textit{Bottom left:} Approximate Bayesian computation plot of the DTC. \textit{Bottom right:} Approximate Bayesian computation plot of the RFC. All plots are for the source A1795Sy1 where the classifiers have been trained on the resid data.}
\label{fig:dtc_rfc_comp}
\end{figure*}

\item {\bf Method comparison:} 
We find that the ML methods return tighter constraints than those from the $\chi^2$ statistic. In particular, for sources where only a small number of counts is available (e.g.~the quasar behind A1795 and the quasar SDSS J130001.48+275120.6 behind Coma) the advantage of the ML methods is significant. For those sources the $\chi^2$ statistic does not return bounds at a 95\% C.L., whereas the ML methods are able to constrain ALPs very well at that confidence level. Out of the three ML methods, the single coupling method provides the best constraints. However, these best bounds often arise from initially worse performing classifiers as discussed previously. We find that the ApBC method is able to resolve that problem and gives more consistent bounds across different classifiers and data products. Furthermore, in cases where the test statistic of the real data is very large (e.g.~for the resid data of A1367) and the single coupling method does not return bounds, we are able to report bounds with the ApBC method. This however is only true if the test statistic of the real data is large but not maximal. Therefore, the ApBC method cannot return constraints for the up data from all quasars either. Nevertheless we argue that due to these advantages the ApBC method should be used preferably. The multiclass classification method gave very good constraints ($g_{a\gamma\gamma} \lesssim 0.6 \times 10^{-12} \, \mbox{GeV}^{-1}$) for the deep neural network.

\end{enumerate}

\begin{table}
\centering
\caption{Constraints on the ALP-photon coupling in units of $10^{-12}\, \mbox{GeV}^{-1}$ for various methods, classifiers and data products from the source A1795Sy1.}
\label{tab:bounds_A1795Sy1}
\begin{tabular}{|c| c c c|}
\hline
\multicolumn{4}{|c|}{A1795Sy1} \\
\hline
\hline
 Method & Classifier & Data product & Constraint \\
\hline
 $\chi^2$-statistic & & & 0.9 \\
 \hline
 \multirow{19}{*}{Single ML} & DTC & resid & 0.5 \\
 \cline{2-4}
 & RFC & up-resid & 0.5 \\
 \cline{2-4}
 & DTC & up & 0.5 \\
 \cline{2-4}
 & DTC & up-resid & 0.6 \\
 \cline{2-4}
 & SVM & resid & 0.8 \\
 \cline{2-4}
 & QDA & up-resid & 0.8 \\
 \cline{2-4}
 & GNB & up-resid & 0.8 \\
  \cline{2-4}
 & ABC & up-resid & 0.8 \\
  \cline{2-4}
 & GNB & up & 0.8 \\
   \cline{2-4}
 & RFC & up & 0.8 \\
   \cline{2-4}
 & ABC & up & 0.8 \\
   \cline{2-4}
 & QDA & resid & 0.9 \\
   \cline{2-4}
 & GNB & resid & 0.9 \\
   \cline{2-4}
 & RFC & resid & 0.9 \\
   \cline{2-4}
 & ABC & resid & 0.9 \\
   \cline{2-4}
 & QDA & up & 0.9 \\
   \cline{1-4}
 \multirow{21}{*}{ApBC} & ABC & up-resid & 0.8 \\
   \cline{2-4}
 & QDA & up & 0.8 \\
    \cline{2-4}
 & GNB & up & 0.8 \\
    \cline{2-4}
 & RFC & up & 0.8 \\
    \cline{2-4}
 & ABC & up & 0.8 \\
    \cline{2-4}
 & QDA & resid & 0.9 \\
     \cline{2-4}
 & GNB & resid & 0.9 \\
     \cline{2-4}
 & RFC & resid & 0.9 \\
     \cline{2-4}
 & ABC & resid & 0.9 \\
     \cline{2-4}
 & QDA & up-resid & 0.9 \\
      \cline{2-4}
 & GNB & up-resid & 0.9 \\
      \cline{2-4}
 & DTC & up-resid & 0.9 \\
      \cline{2-4}
 & RFC & up-resid & 0.9 \\
      \cline{2-4}
 & DTC & up & 0.9 \\
       \cline{2-4}
 & SVM & up & 0.9 \\
       \cline{2-4}
 & SVM & up-resid & 1.0 \\
       \cline{2-4}
 & DTC & resid & 1.1 \\
       \cline{2-4}
 & SVM & resid & 1.4 \\
 \hline
 \multirow{6}{*}{Multiclass} & DNN & up-resid & 0.6 \\
 \cline{2-4}
 & DNN & up & 0.6 \\
 \cline{2-4}
 & DNN & resid & 0.8 \\
 \cline{2-4}
 & QDA & up-resid & 0.8 \\
 \cline{2-4}
 & QDA & up & 0.9 \\
 \cline{2-4}
 & QDA & resid & 1.0\\

 \hline
\end{tabular}
\end{table}

\begin{table}
\centering
\caption{Constraints on the ALP-photon coupling in units of $10^{-12}\, \mbox{GeV}^{-1}$ for various methods, classifiers and data products from the source A1367.}
\label{tab:bounds_A1367}
\begin{tabular}{|c| c c c|}
\hline
\multicolumn{4}{|c|}{A1367} \\
\hline
\hline
 Method & Classifier & Data product & Constraint \\
\hline
 $\chi^2$-statistic & & & 2.0 \\
 \hline
 \multirow{8}{*}{Single ML} & RFC & up-resid & 1.2 \\
 \cline{2-4}
 & SVM & up-resid & 1.2 \\
 \cline{2-4}
 & QDA & up & 1.2 \\
 \cline{2-4}
 & GNB & up & 1.3 \\
 \cline{2-4}
 & DTC & up & 1.6 \\
 \cline{2-4}
 & RFC & up & 1.6 \\
 \cline{2-4}
 & SVM & up & 1.6 \\
 \cline{2-4}
 & GNB & up-resid & 1.8 \\
 \hline
 \multirow{14}{*}{ApBC} & QDA & up & 1.4 \\
 \cline{2-4}
 & GNB & up & 1.6 \\
 \cline{2-4}
 & RFC & up & 1.6 \\
 \cline{2-4}
 & SVM & up & 1.6 \\
 \cline{2-4}
 & RFC & up-resid & 1.7 \\
 \cline{2-4}
 & GNB & up-resid & 1.8 \\
 \cline{2-4}
 & ABC & up-resid & 1.8 \\
 \cline{2-4}
 & SVM & up-resid & 1.8 \\
 \cline{2-4}
 & DTC & up & 1.8 \\
 \cline{2-4}
 & QDA & resid & 1.9 \\
 \cline{2-4}
 & RFC & resid & 1.9 \\
 \cline{2-4}
 & ABC & resid & 1.9 \\
 \cline{2-4}
 & QDA & up-resid & 1.9 \\
 \cline{2-4}
 & ABC & up & 1.9 \\
 \hline
 \multirow{2}{*}{Multiclass} & QDA & up & 1.2 \\
 \cline{2-4}
 & DNN & up & 1.4 \\

 \hline
\end{tabular}
\end{table}

\begin{table}
\centering
\caption{Constraints on the ALP-photon coupling in units of $10^{-12}\, \mbox{GeV}^{-1}$ for various methods, classifiers and data products from the source A1795Quasar.}
\label{tab:bounds_A1795Quasar}
\begin{tabular}{|c| c c c|}
\hline
\multicolumn{4}{|c|}{A1795Quasar} \\
\hline
\hline
 Method & Classifier & Data product & Constraint \\
\hline
 $\chi^2$-statistic & & & 1.0 (84\% C.L.)  \\
 \hline
 \multirow{5}{*}{Single ML} & SVM & resid & 0.4 \\
 \cline{2-4}
 & DTC & up-resid & 0.5 \\
 \cline{2-4}
 & GNB & up-resid & 0.7 \\
 \cline{2-4}
 & QDA & resid & 0.9 \\
 \cline{2-4}
 & RFC & resid & 1.0 \\
 \hline
 \multirow{5}{*}{ApBC} & SVM & resid & 0.6 \\
 \cline{2-4}
 & QDA & resid & 1.6\\
 \cline{2-4}
 & DTC & up-resid & 1.8 \\
 \cline{2-4}
 & RFC & resid & 1.9\\
 \cline{2-4}
 & GNB & up-resid & 1.9\\
 \hline
 \multirow{3}{*}{Multiclass} & DNN & resid & 0.6 \\
 \cline{2-4}
 & QDA & up-resid & 0.8 \\
 \cline{2-4}
 & QDA & resid & 1.0 \\
 \hline
\end{tabular}
\end{table}

\begin{table}
\centering
\caption{Constraints on the ALP-photon coupling in units of $10^{-12}\, \mbox{GeV}^{-1}$ for various methods, classifiers and data products from the source Coma1.}
\label{tab:bounds_Coma1}
\begin{tabular}{|c| c c c|}
\hline
\multicolumn{4}{|c|}{Coma1} \\
\hline
\hline
 Method & Classifier & Data product & Constraint \\
\hline
 $\chi^2$-statistic & & & 2.5 \\
 \hline
 \multirow{6}{*}{Single ML} & SVM & resid & 1.5 \\
 \cline{2-4}
 & DTC & up-resid & 1.7 \\
 \cline{2-4}
 & RFC & up-resid & 1.9 \\
 \cline{2-4}
 & ABC & up-resid & 2.2 \\
 \cline{2-4}
 & SVM & up-resid & 2.5 \\
 \cline{2-4}
 & GNB & up-resid & 2.7 \\
 \hline
 \multirow{6}{*}{ApBC} & SVM & resid & 2.1 \\
 \cline{2-4}
 & RFC & up-resid & 2.8 \\
 \cline{2-4}
 & QDA & up-resid & 2.9 \\
 \cline{2-4}
 & GNB & up-resid & 2.9 \\
 \cline{2-4}
 & ABC & up-resid & 2.9 \\
 \cline{2-4}
 & SVM & up-resid & 2.9 \\
 \hline
 \multirow{4}{*}{Multiclass} & QDA & up-resid & 1.7 \\
 \cline{2-4}
 & DNN & resid & 1.8 \\
 \cline{2-4}
 & DNN & up-resid & 2.0 \\
 \cline{2-4}
 & QDA & resid & 2.2 \\
 \hline
\end{tabular}
\end{table}

\begin{table}
\centering
\caption{Constraints on the ALP-photon coupling in units of $10^{-12}\, \mbox{GeV}^{-1}$ for various methods, classifiers and data products from the source Coma2.}
\label{tab:bounds_Coma2}
\begin{tabular}{|c| c c c|}
\hline
\multicolumn{4}{|c|}{Coma2} \\
\hline
\hline
 Method & Classifier & Data product & Constraint \\
\hline
 $\chi^2$-statistic & & & 3.0 (90\% C.L.) \\
 \hline
 \multirow{7}{*}{Single ML} & ABC & resid & 1.3 \\
 \cline{2-4}
 & DTC & up-resid & 1.7 \\
 \cline{2-4}
 & QDA & resid & 1.8 \\
 \cline{2-4}
 & RFC & resid & 1.9 \\
 \cline{2-4}
 & SVM & resid & 1.9 \\
 \cline{2-4}
 & QDA & up-resid & 2.9 \\
 \cline{2-4}
 & ABC & up-resid & 3.0 \\
 \hline
 \multirow{6}{*}{ApBC} & QDA & resid & 2.8 \\
 \cline{2-4}
 & ABC & resid & 2.8 \\
 \cline{2-4}
 & GNB & resid & 2.9 \\
 \cline{2-4}
 & RFC & resid & 2.9 \\
 \cline{2-4}
 & SVM & resid & 2.9 \\
 \cline{2-4}
 & QDA & up-resid & 2.9 \\
 \hline
 \multirow{3}{*}{Multiclass} & QDA & resid & 1.8 \\
 \cline{2-4}
 & DNN & resid & 2.0 \\
 \cline{2-4}
 & QDA & up-resid & 2.9 \\
 \hline
\end{tabular}
\end{table}

\subsection{Bounds with restricted energy range}
A natural question is how important individual spectral feature are for the ML bounds. Here we are interested in analysing the effect a restriction of the energy range has on our results.
Beyond the standard data analysis question of feature importance, this feature restriction is relevant for the on-going all-sky survey \textit{eROSITA}~\citep{erosita}. Here the energy range of the expected point-source spectra will be restricted as effective area $\times$ field of view is largest for energies smaller than $2.5 \, \mbox{keV}.$

To estimate how our ALP bounds are affected by this restricted energy range we have checked whether it is still possible to constrain ALPs when we restrict our analysis to this energy range below $2.5 \, \mbox{keV}.$ We list the bounds for all sources and data products using spectra with this energy range in Table~\ref{tab:erosita}. Even though the resulting bounds are not as tight as when using the complete spectrum, they demonstrate that we can obtain good bounds on the ALP-photon coupling in that energy range.

\begin{table*}
\centering
\caption{Constraints on the ALP-photon coupling in units of $10^{-12}\, \mbox{GeV}^{-1}$ for all sources and data products using only the low energy part of the spectra ($E<2.5 \, \mbox{keV}$). We have used the QDA classifiers and approximate Bayesian computation.}
\label{tab:erosita}
\begin{tabular}{|c||c|c|c|c|c|}
 \hline
 Data product & A1795Sy1 & A1367 & A1795Quasar & Coma1 & Coma2 \\
 \hline 
 \hline
 resid & 1.3 & 1.8 & 2.0 & - & 2.9 \\
 \hline
 up-resid & 1.1 & 1.8 & - & 2.9 & 3.0 \\
 \hline
 up & 1.1 & 1.3 & - & - & - \\
 \hline
\end{tabular}
\end{table*}

\section{Conclusion and Outlook}\label{sec:conclusion}

In this work we were able to constrain the coupling constant between ALPs and photons to $g_{a\gamma\gamma} \lesssim 0.6 \times 10^{-12} \, \mbox{GeV}^{-1}$ (95\% C.L.). These are the best bounds on ALPs for the observations that we have used and at the same level as current state-of-the-art bounds~\citep{grating}.

We have applied for the first time 3D magnetic field simulations of galaxy clusters in order to place bounds in the X-ray regime. Throughout all sources the 3D model gave tighter bounds than previously used 1D simulations. As we have seen, this is due to a more precise (less conservative) normalization of the magnetic field. However, even the upscaled 1D model did not match the bounds from the 3D model. We suspect that this is due to an inherent difference in the models where the 3D one leads to bigger amplitudes in the magnetic field strength. A confirmation of this is left for the future.

Furthermore, we present three different ML methods which are able to improve the constraints, especially for sources with a poor spectral quality. For the first time we use approximate Bayesian computation in order to constrain ALPs, which provides more consistent bounds across classifiers. Across classifiers, the best bounds we find are in the multiclass classification when applying a deep neural network and in the approximate Bayesian computation method when using a support vector machine.

We also find that restricting the energy range of our spectra only results in slightly worse bounds which is of high relevance for the on-going \textit{eROSITA} mission.

Given this improvement when using ML-based methods to search for ALPs it would be of great interest to revisit the expected bounds for the future X-ray mission \textit{ATHENA} which will have an outstanding energy resolution combined with longer observations times~\citep{athena}. Also, X-ray polarimeters such as \textit{IXPE} may provide tighter bounds~\citep{polari}.

Additionally, given the sensitivity to the magnetic field model we find, an improvement in the modelling and the observational constraints on magnetic fields in galaxy clusters seems very relevant for ALP searches. We hope that future radio observations such as with the \textit{Square Kilometre Array} will improve upon these uncertainties in the magnetic fields and vitally provide magnetic field information for a large number of clusters~\citep{ska}.
In conclusion, magnetic field estimates as well as the resolution of the spectra of point sources in the X-ray regime will significantly improve in the future. To optimally utilize this new data for constraining ALPs, developing sophisticated techniques such as machine learning is important.

 \section*{Acknowledgements}
 
 Funded by the Deutsche Forschungsgemeinschaft (DFG, German Research Foundation) under Germany´s Excellence Strategy – EXC-2094 – 390783311. FCD is supported by Stephen Hawking Fellowship EP/T01668X/1 and STFC grant ST/P001246/1.
 
 \section*{Data Availability}
 
 The data underlying this article are available in the Chandra Data Archive at \url{https://cda.harvard.edu/chaser/}.
 
\bibliographystyle{mnras}
\bibliography{axion-ml}

\appendix

\section{Observation IDs} \label{app:obs_ids}
In Table~\ref{tab:obs_ids} we list all incorporated \textit{Chandra} observation IDs.
\begin{table*}
\centering
\caption{\textit{Chandra} observation IDs of the galaxy clusters.}
\label{tab:obs_ids}
\begin{tabular}{|c|c|}
 \hline
 Cluster & IDs \\
 \hline
  \hline
 \multirow{9}{*}{A1795} & 493, 494, 3666, 5286, 5287, 5288, 5289, 5290, 6159, 6160, \\
 & 6161, 6162, 6163, 10898, 10900, 12026, 12027, 12028, 12029, 13106 \\
 & 13107, 13108, 13109, 13110, 13111, 13112, 13412, 13413, 13414, 13415 \\
 & 13416, 13417, 14268, 14269, 14270, 14271, 14272, 14273, 14274, 14275 \\
 & 15485, 15486, 15487, 15488, 15491, 15492, 16433, 16434, 16436, 16437 \\
 & 16438, 16439, 16465, 16467, 16468, 16469, 16471, 16472, 17397, 17398 \\
 & 17399, 17401, 17402, 17403, 17404, 17405, 17406, 17407, 17408, 17410 \\
 & 17411, 17683, 17684, 17685, 17686, 18423, 18424, 18425, 18426, 18427 \\
 & 18429, 18430, 18431, 18432, 18433, 18434, 18435, 18436, 18438, 18439\\
 \hline
 A1367 & 514, 4916 \\
 \hline
 \multirow{2}{*}{Coma} & 555, 556, 1086, 1112, 1113, 1114, 9714,     10672, 13993, 13994 \\
 & 13995, 13996, 14406, 14410, 14411, 14415\\
 \hline
\end{tabular}
\end{table*}

\section{Hyperparameters}\label{app:hyperparameters}

This appendix summarizes the best hyperparameters which we found in the grid search and used to train the classifiers for the single coupling ML and ApBC method. Here, we only mention hyperparameters that have been optimized in the grid search, i.e.~for all other hyperparameters we have used the default values of \textit{scikit-learn} (version 0.23.2).

\subsection*{A1795Sy1}

Coupling used for the grid search: $g_{a\gamma\gamma}=10^{-12}\, \mbox{GeV}^{-1}$. We find the following best hyperparameters:
\begin{enumerate}
    \item resid:
    \begin{description}
        \item DTC(max\_depth=100, min\_samples\_split=100,\\ min\_samples\_leaf=1)
        \item RFC(n\_estimators=500, max\_depth=100, min\_samples\_split=10,\\ min\_samples\_leaf=1)
        \item ABC(n\_estimators=500, learning\_rate=1.0)
        \item SVM(C=100.0)
    \end{description}
    \item up:
    \begin{description}
        \item DTC(max\_depth=100, min\_samples\_split=100,\\ min\_samples\_leaf=1)
        \item RFC(n\_estimators=500, max\_depth=100, min\_samples\_split=5,\\ min\_samples\_leaf=1)
        \item ABC(n\_estimators=500, learning\_rate=1.0)
        \item SVM(C=100.0)
    \end{description}
\end{enumerate}

\subsection*{A1367}
We use the coupling $g_{a\gamma\gamma}= 1.5 \times 10^{-12}\, \mbox{GeV}^{-1}$ for the grid search. Below we list the best hyperparameters:
\begin{enumerate}
    \item resid:
    \begin{description}
        \item DTC(max\_depth=800, min\_samples\_split=100,\\ min\_samples\_leaf=35)
        \item RFC(n\_estimators=150, max\_depth=100, min\_samples\_split=10,\\ min\_samples\_leaf=2)
        \item ABC(n\_estimators=150, learning\_rate=1.0)
        \item SVM(C=1.0)
    \end{description}
    \item up:
    \begin{description}
        \item DTC(max\_depth=550, min\_samples\_split=150,\\ min\_samples\_leaf=80)
        \item RFC(n\_estimators=150, max\_depth=100, min\_samples\_split=2,\\ min\_samples\_leaf=1)
        \item ABC(n\_estimators=150, learning\_rate=0.5)
        \item SVM(C=100.0)
    \end{description}
\end{enumerate}

\subsection*{A1795Quasar}
We use the coupling $g_{a\gamma\gamma}= 1.5 \times 10^{-12}\, \mbox{GeV}^{-1}$ for the grid search. Below we list the best hyperparameters:
\begin{enumerate}
    \item resid:
    \begin{description}
        \item DTC(max\_depth=None, min\_samples\_split=100,\\ min\_samples\_leaf=10)
        \item RFC(n\_estimators=150, max\_depth=None, min\_samples\_split=2,\\ min\_samples\_leaf=2)
        \item ABC(n\_estimators=150, learning\_rate=0.9)
        \item SVM(C=100.0)
    \end{description}
    \item up:
    \begin{description}
        \item DTC(max\_depth=None, min\_samples\_split=2,\\ min\_samples\_leaf=500)
        \item RFC(n\_estimators=100, max\_depth=100, min\_samples\_split=2,\\ min\_samples\_leaf=2)
        \item ABC(n\_estimators=150, learning\_rate=0.8)
        \item SVM(C=100.0)
    \end{description}
\end{enumerate}

\subsection*{Coma1}
We use the coupling $g_{a\gamma\gamma}= 2 \times 10^{-12}\, \mbox{GeV}^{-1}$ for the grid search. Below we list the best hyperparameters:
\begin{enumerate}
    \item resid:
    \begin{description}
        \item DTC(max\_depth=None, min\_samples\_split=500,\\ min\_samples\_leaf=10)
        \item RFC(n\_estimators=150, max\_depth=None,\\ min\_samples\_split=5, min\_samples\_leaf=10)
        \item ABC(n\_estimators=150, learning\_rate=0.9)
        \item SVM(C=100.0)
    \end{description}
    \item up:
    \begin{description}
        \item DTC(max\_depth=10, min\_samples\_split=100,\\ min\_samples\_leaf=5)
        \item RFC(n\_estimators=150, max\_depth=500, min\_samples\_split=10,\\ min\_samples\_leaf=2)
        \item ABC(n\_estimators=150, learning\_rate=0.7)
        \item SVM(C=100.0)
    \end{description}
\end{enumerate}

\subsection*{Coma2}
We use the coupling $g_{a\gamma\gamma}= 2.7 \times 10^{-12}\, \mbox{GeV}^{-1}$ for the grid search. Below we list the best hyperparameters:
\begin{enumerate}
    \item resid:
    \begin{description}
        \item DTC(max\_depth=10, min\_samples\_split=500,\\ min\_samples\_leaf=1)
        \item RFC(n\_estimators=150, max\_depth=None, min\_samples\_split=100,\\ min\_samples\_leaf=2)
        \item ABC(n\_estimators=150, learning\_rate=0.9)
        \item SVM(C=100.0)
    \end{description}
    \item up:
    \begin{description}
        \item DTC(max\_depth=None, min\_samples\_split=500,\\ min\_samples\_leaf=1)
        \item RFC(n\_estimators=150, max\_depth=100, min\_samples\_split=2,\\ min\_samples\_leaf=5)
        \item ABC(n\_estimators=150, learning\_rate=0.9)
        \item SVM(C=100.0)
    \end{description}
\end{enumerate}

\bsp	
\label{lastpage}
\end{document}